\documentclass[acmsmall]{acmart}

\usepackage[utf8]{inputenc}
\usepackage[T1]{fontenc}
\usepackage{hyphenat}
\usepackage{xspace}
\usepackage{amsmath}
\usepackage{amsfonts}
\usepackage{hyperref}
\usepackage{url}
\usepackage{multirow}
\usepackage{makecell}
\usepackage{caption}
\usepackage{minibox}
\usepackage{bbm}
\usepackage{graphicx}
\usepackage{balance}
\usepackage{mathtools}
\usepackage{color}
\usepackage{marvosym}
\usepackage{ifthen}
\usepackage{textcomp}
\usepackage{enumitem}
\usepackage{verbatim}
\usepackage{algorithm}
\usepackage{algorithmic}
\usepackage{numprint}
\usepackage{balance}

\usepackage{subcaption}

\usepackage{amsthm}
\theoremstyle{plain}

\newcommand{\chatoDisplayMode}[1]{#1}

\definecolor{MyRed}{rgb}{0.6,0.0,0.0} 
\definecolor{MyBlack}{rgb}{0.1,0.1,0.1} 
\newcommand{\inred}[1]{{\color{MyRed}\sf\textbf{\textsc{#1}}}}
\newcommand{\frameit}[2]{
  \begin{center}
  {\color{MyRed}
  \framebox[.9\columnwidth][l]{
    \begin{minipage}{.85\columnwidth}
    \inred{#1}: {\sf\color{MyBlack}#2}
    \end{minipage}
  }\\
  }
  \end{center}
}

\newcommand{\note}[2][]{\chatoDisplayMode{\def\@tmpsig{#1}\frameit{{\Pointinghand} Note}{#2\ifx \@tmpsig \@empty \else \mbox{ --\em #1}\fi}}}
\newcommand{\todo}[2][]{\chatoDisplayMode{\def\@tmpsig{#1}\frameit{{\Writinghand} To-do}{#2\ifx \@tmpsig \@empty \else \mbox{ --\em #1}\fi}}}

\newcommand{\rev}[1]{\textcolor{black}{#1}}
\newcommand{\minor}[1]{\textcolor{black}{#1}}

\newcommand{\abbrevStyle}[1]{#1}

\newcommand{\ie}{\abbrevStyle{i.e.}\xspace}
\newcommand{\eg}{\abbrevStyle{e.g.}\xspace}
\newcommand{\cf}{\abbrevStyle{cf.}\xspace}

\newcommand{\vs}{\abbrevStyle{vs.}\xspace}
\newcommand{\etc}{\abbrevStyle{etc.}\xspace}

\newcommand{\Secref}[1]{Sec.~\ref{#1}}
\newcommand{\Eqnref}[1]{Eq.~\ref{#1}}

\newcommand{\Tabref}[1]{Table~\ref{#1}}
\newcommand{\Figref}[1]{Fig.~\ref{#1}}

\newcommand{\xhdrIt}[1]{\vspace{1.5mm}\noindent{\emph{#1.}}}

\newcommand{\textcite}[1]{\citeauthor{#1} \shortcite{#1}}

\newcommand{\hide}[1]{}

\makeatletter
\newcommand{\iffont}[2]{\ifthenelse{\equal{\f@family}{#1}}{#2}{}}
\makeatother

\iffont{ptm}{
  \usepackage{mathptmx}

  \DeclareSymbolFont{greek}{OML}{cmm}{m}{n}
  \DeclareMathSymbol{\alpha}{\mathalpha}{greek}{"0B}
  \DeclareMathSymbol{\beta}{\mathalpha}{greek}{"0C}
  \DeclareMathSymbol{\gamma}{\mathalpha}{greek}{"0D}
  \DeclareMathSymbol{\delta}{\mathalpha}{greek}{"0E}
  \DeclareMathSymbol{\epsilon}{\mathalpha}{greek}{"0F}
  \DeclareMathSymbol{\zeta}{\mathalpha}{greek}{"10}
  \DeclareMathSymbol{\eta}{\mathalpha}{greek}{"11}
  \DeclareMathSymbol{\theta}{\mathalpha}{greek}{"12}
  \DeclareMathSymbol{\iota}{\mathalpha}{greek}{"13}
  \DeclareMathSymbol{\kappa}{\mathalpha}{greek}{"14}
  \DeclareMathSymbol{\lambda}{\mathalpha}{greek}{"15}
  \DeclareMathSymbol{\mu}{\mathalpha}{greek}{"16}
  \DeclareMathSymbol{\nu}{\mathalpha}{greek}{"17}
  \DeclareMathSymbol{\xi}{\mathalpha}{greek}{"18}
  \DeclareMathSymbol{\pi}{\mathalpha}{greek}{"19}
  \DeclareMathSymbol{\rho}{\mathalpha}{greek}{"1A}
  \DeclareMathSymbol{\sigma}{\mathalpha}{greek}{"1B}
  \DeclareMathSymbol{\tau}{\mathalpha}{greek}{"1C}
  \DeclareMathSymbol{\upsilon}{\mathalpha}{greek}{"1D}
  \DeclareMathSymbol{\phi}{\mathalpha}{greek}{"1E}
  \DeclareMathSymbol{\chi}{\mathalpha}{greek}{"1F}
  \DeclareMathSymbol{\psi}{\mathalpha}{greek}{"20}
  \DeclareMathSymbol{\omega}{\mathalpha}{greek}{"21}
  \DeclareMathSymbol{\varepsilon}{\mathalpha}{greek}{"22}
  \DeclareMathSymbol{\vartheta}{\mathalpha}{greek}{"23}
  \DeclareMathSymbol{\varpi}{\mathalpha}{greek}{"24}
  \DeclareMathSymbol{\varrho}{\mathalpha}{greek}{"25}
  \DeclareMathSymbol{\varsigma}{\mathalpha}{greek}{"26}
  \DeclareMathSymbol{\varphi}{\mathalpha}{greek}{"27}
  \DeclareSymbolFont{otone}{OT1}{cmr}{m}{n}
  \DeclareMathSymbol{\Gamma}{\mathalpha}{otone}{0}
  \DeclareMathSymbol{\Delta}{\mathalpha}{otone}{1}
  \DeclareMathSymbol{\Theta}{\mathalpha}{otone}{2}
  \DeclareMathSymbol{\Lambda}{\mathalpha}{otone}{3}
  \DeclareMathSymbol{\Xi}{\mathalpha}{otone}{4}
  \DeclareMathSymbol{\Pi}{\mathalpha}{otone}{5}
  \DeclareMathSymbol{\Sigma}{\mathalpha}{otone}{6}
  \DeclareMathSymbol{\Upsilon}{\mathalpha}{otone}{7}
  \DeclareMathSymbol{\Phi}{\mathalpha}{otone}{8}
  \DeclareMathSymbol{\Psi}{\mathalpha}{otone}{9}
  \DeclareMathSymbol{\Omega}{\mathalpha}{otone}{10}
  \DeclareSymbolFont{syms}{OML}{cmm}{m}{it}
  \DeclareMathSymbol{\partial}{\mathord}{syms}{"40}
  \DeclareMathAlphabet{\mathbold}{OML}{cmm}{b}{it}
  \DeclareSymbolFont{largesymbols}{OMX}{cmex}{m}{n}

}
\usepackage{xcolor}
\usepackage{flushend}
\usepackage{csquotes}

\newcommand{\KL}{D_{\text{KL}}}
\newcommand{\pmfr}{p_{\text{M}}}
\newcommand{\pt}{p_{\text{T}}}

\AtBeginDocument{%
  \providecommand\BibTeX{{%
    \normalfont B\kern-0.5em{\scshape i\kern-0.25em b}\kern-0.8em\TeX}}}

\begin{document}

\title{Biased \minor{Bytes}: On the Validity of Estimating Food Consumption from \rev{Digital Traces}}

\author{Kristina Gligori\'c}
\affiliation{%
  \institution{EPFL}
  \country{Lausanne, Switzerland}
}
\email{kristina.gligoric@epfl.ch}

\author{Irena Đorđevi\'c}
\authornote{Work done during an internship at EPFL.}
\affiliation{%
  \institution{University of Niš}
  \country{Niš, Serbia}
}
\email{irenadj@elfak.rs}

\author{Robert West}
\affiliation{%
  \institution{EPFL}
  \country{Lausanne, Switzerland}
}
\email{robert.west@epfl.ch}

\renewcommand{\shortauthors}{Gligori\'c et al.}

\begin{abstract}
Given that measuring food consumption at a population scale is a challenging task, researchers have begun to explore digital traces (\eg, from social media or from food-tracking applications) as potential proxies. However, it remains unclear to what extent digital traces reflect real food consumption. The present study aims to bridge this gap by quantifying the link between dietary behaviors as captured via social media (Twitter) \vs\ a food-tracking application (MyFoodRepo).
We focus on the case of Switzerland and contrast images of foods collected through the two platforms, by designing and deploying a novel crowdsourcing framework for estimating biases with respect to nutritional properties and appearance.
We find that the food type distributions in social media \vs\ food tracking diverge; \eg, bread is 2.5 times more frequent among consumed and tracked foods than on Twitter, whereas cake is 12 times more frequent on Twitter. Controlling for the different food type distributions, we contrast consumed and tracked foods of a given type with foods shared on Twitter. Across food types, food posted on Twitter is perceived as tastier, more caloric, less healthy, less likely to have been consumed at home, more complex, and larger-portioned, compared to consumed and tracked foods.
The fact that there is a divergence between food consumption as measured via the two platforms implies that at least one of the two is not a faithful representation of the true food consumption in the general Swiss population. Thus, researchers should be attentive and aim to establish evidence of validity before using digital traces as a proxy for the true food consumption of a general population. We conclude by discussing the potential sources of these biases and their implications, outlining pitfalls and threats to validity, and proposing actionable ways for overcoming them.


\end{abstract}

\setcopyright{acmlicensed}
\acmJournal{PACMHCI}
\acmYear{2022} \acmVolume{6} \acmNumber{CSCW2} \acmArticle{497} \acmMonth{11} \acmPrice{15.00}\acmDOI{10.1145/3555660}

\ccsdesc[500]{Human-centered computing~Empirical studies in collaborative and social computing}
\ccsdesc[500]{Human-centered computing~Social media}
\ccsdesc[500]{Applied computing~Health informatics}
\ccsdesc[500]{Information systems~Data mining}
\ccsdesc[500]{Information systems~Social networking sites}
\ccsdesc[500]{General and reference~Empirical studies}
\ccsdesc[500]{General and reference~Validation}

\keywords{biases, validity, social media, food, images, Twitter, crowdsourcing}

\maketitle

\section{Introduction}
\label{sec:intro}


Diets determine health. Eating healthy helps prevent malnutrition as well as a range of diseases and conditions, including diabetes, heart disease, stroke, and cancer \cite{habib2010burden}. In order to be able to improve diets, researchers and stakeholders need to know what foods people consume, but monitoring diets at a population scale is challenging. Traditionally, nutritional studies rely on survey\hyp based methods \cite{christakis2007spread} employing questionnaires \cite{wouters2010peer} and personal food journals \cite{schroeder2017supporting,barriers_negative2015,food_journal2015}, which are prone to biases, most notably social and cognitive biases, such as false recall and social desirability bias \cite{bowling2005mode}. 
Traditional methods are also costly to organize. Researchers and practitioners might not have access to extensive surveying, and it might be hard to collect reliable statistics---even though a large and ever-growing \cite{poushter2016smartphone} portion of the population has access to advanced technology including smartphones with Internet access \cite{blumenstock2015predicting}.

In light of the challenges of traditional methods on the one hand,
and the opportunities afforded by widespread Internet access on the other hand, 
there is great promise in using passively collected digital data to estimate food consumption. 
Digital datasets are unmatched in terms of scale and immediacy \cite{salganik2019bit,mejova2015twitter}, 
do not rely on self\hyp reports, and 
do not suffer from biases typical of traditional methods. 
Given this potential, researchers have been developing and applying their expertise to studying diets via passively collected digital data, whose tremendous potential for providing insights into food consumption has been showcased numerous times \cite{abbar2015you,saura2020gaining,chung2015more}.


\begin{figure}[b!]
    \centering
    \includegraphics[width = \columnwidth]{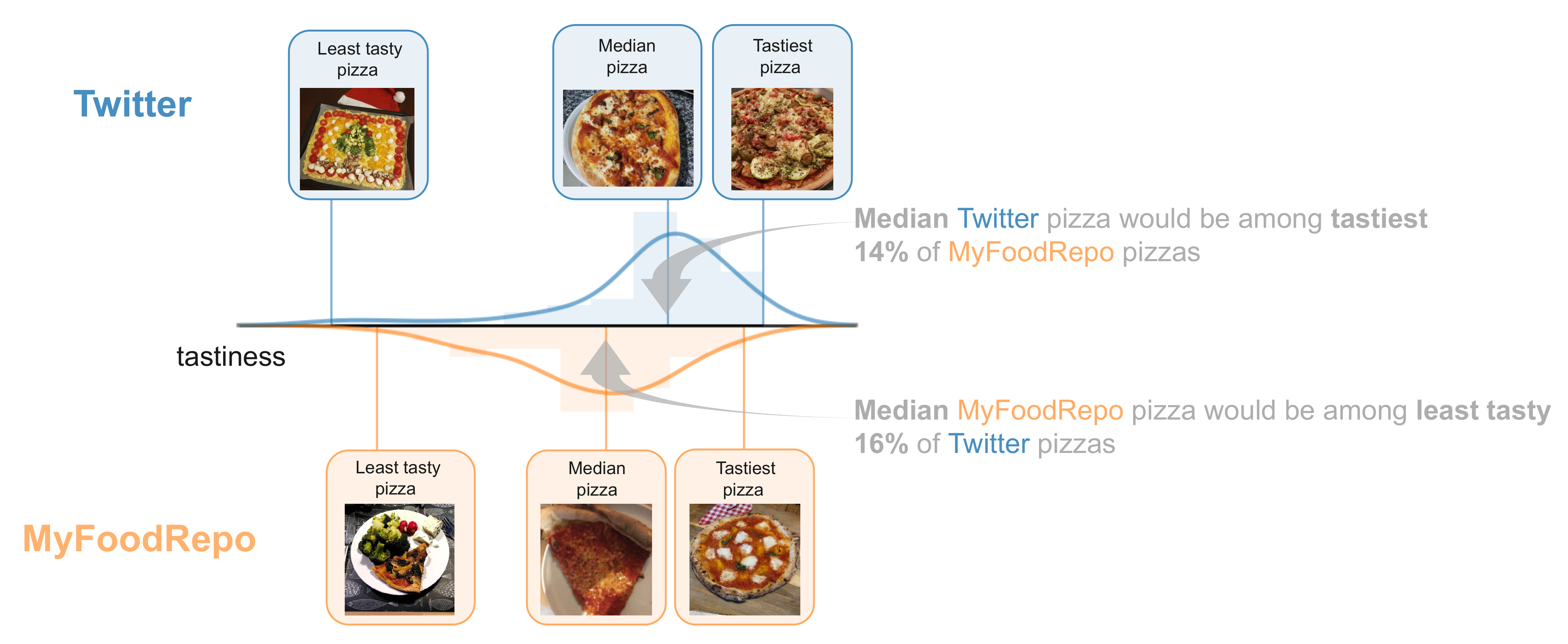}
    \caption{\textbf{Illustration of bias in perceived tastiness.} Perceived tastiness of tweeted food (\textit{top}) \vs\ actually consumed \rev{and tracked} food (\textit{bottom}) of type ``pizza''. Histograms summarize tastiness scores estimated in our crowdsourcing framework.
    As illustrated, tweeted pizzas are perceived as considerably tastier than actually consumed \rev{and tracked} pizzas.
    }
    \label{fig:diagram}
\end{figure}

The promises of Web and social media data notwithstanding, important methodological questions remain: Are researchers measuring what they aim to measure? Do digital traces reflect actual food consumption? Do effects estimated from online signals hold in the offline world? Are predictive models trained on online signals accurate in the offline world? In other words, the validity of studying diets with digital data remains opaque. 

Online data is not primarily collected with scientific studies in mind and is therefore sometimes referred to as ``found data'' \cite{salganik2019bit}. Found data overcomes several of the biases typical of traditional methods, but may introduce new biases that threaten validity in their own ways \cite{lazer2021meaningful,olteanu2019social,wagner2021measuring}. Despite their potential, error\hyp prone and unreliable data and methods may do more harm than good if handled without the required caution \cite{corbett2017algorithmic}. As our community increasingly relies on large-scale digital data sources, methods that offer insights into the validity of new measures thus become increasingly necessary \cite{lazer2021meaningful}.



\subsection{Research questions}

In this spirit, our overall research question asks: \emph{Is social media a biased or a truthful mirror of actual food consumption, \rev{as measured via food tracking}?} We focus on those dietary aspects in particular that researchers frequently study with social media: food type \cite{de2016characterizing}, nutritional properties \cite{abbar2015you}, and appearance and subjective perception~\cite{mejova2015foodporn}.
In order to establish a link between online and offline dietary behaviors at a population scale, we study images of food that people consumed \rev{and tracked} via a food-tracking application, and contrast them with images of food posted on Twitter, addressing the following specific research questions:

\begin{enumerate}
\item[\textbf{RQ1}]
\textit{Bias of food-type distribution:}
To what extent do food images posted on Twitter reflect the types (beef, bread, burger, \etc)\ of actually consumed food, \rev{as measured via food tracking}?
\item[\textbf{RQ2}]
\textit{Biases within food types:}
For a given food type, to what extent do food images posted on Twitter reflect the nutritional  properties, perceived tastiness, and appearance of actually consumed \rev{and tracked} food of that type?
\end{enumerate}

\rev{In order to address RQ1, we investigate whether food images posted on Twitter are a faithful reflection of the types of actually consumed and tracked food or not. \textit{A priori,} one might envision the following potential outcomes:}

\begin{enumerate}

\item[\textbf{\rev{a)}}] \rev{``Food images posted on Twitter are a faithful reflection of the types of actually consumed \rev{and tracked} food, consistent with the demonstrated potential of Twitter to provide insight into dietary choices  \cite{abbar2015you}.''}

\item[\textbf{\rev{b)}}] \rev{``Food images posted on Twitter are not a faithful reflection of the types of actually consumed \rev{and tracked} food, given a variety of challenges in the practices of social media use for research~\cite{olteanu2019social}.''}

\end{enumerate}

\rev{In order to address RQ2, we investigate whether or not food images posted on Twitter are a faithful reflection of actually consumed and tracked food in terms of how healthy, caloric, and tasty\hyp looking the food is. We investigate whether the two sources diverge, and if so, in what direction. \textit{A priori,} one might envision the following potential outcomes:}

\begin{enumerate}

\item[\textbf{\rev{a)}}] \rev{``Food images posted on Twitter are a faithful reflection of actually consumed \rev{and tracked} food in terms of how healthy, caloric, and tasty the food is, consistent with the demonstrated potential of Twitter to provide insight into dietary choices \cite{abbar2015you}.}''

\item[\textbf{\rev{b)}}] ``Tweeted food is healthier, less caloric, and less tasty than consumed \rev{and tracked} food. Social media is increasingly used to promote trendy ingredients and recipes, and clean and healthy eating \cite{chung2017personal}. Social media inspires and connects people interested in healthy eating \cite{mete2019healthy}.''

\item[\textbf{\rev{c)}}] ``Tweeted food is less healthy, more caloric, and tastier than consumed \rev{and tracked} food, consistent with a documented fetishization of food online. Users share appetizing pictures of culinary experiences where exaggerated foods such as sugary desserts dominate over more standard local cuisines \cite{mejova2015foodporn}.''

\end{enumerate}

\subsection{Contributions}

\rev{To the best of our knowledge, ours is the first attempt to investigate the link between online and offline dietary behaviors by studying food images as measured via two platforms, in our case Twitter and the MyFoodRepo\footnote{\url{https://www.myfoodrepo.org/}} food\hyp tracking app}. We design and apply a novel crowdsourcing framework for estimating biases (\Secref{sec:methods}), and we perform a case study of food consumption in Switzerland (\Secref{sec:results}). Controlling for location, period, and food types, we contrast an extensive set of tweeted food images with images of consumed and tracked food.

\rev{We find that food type distributions among social media foods \vs among consumed and tracked foods diverge (RQ1, \Secref{sec:RQ1})}. Controlling for the discrepant food-type distributions by studying food types individually (RQ2, \Secref{sec:RQ2}), we find that Twitter still provides a biased view of food consumption \rev{as measured via food tracking}. Tweeted food is, on average across food types, perceived as more caloric, less healthy, less likely to have been consumed at home, and tastier (example in \Figref{fig:diagram}), compared to actually consumed \rev{and tracked} food. For example, on average across food types, a median-tasty Twitter dish is among the top 26\% tastiest \rev{MyFoodRepo} dishes, and a median-caloric Twitter dish is among the top 34\% most caloric \rev{MyFoodRepo} dishes. While social media traces can be a reasonable proxy of \rev{tracked} consumption for certain foods types (\Figref{f}), we find that, overall, \rev{food shared on social media and consumed \rev{and tracked} food significantly diverge from each other (\Figref{fig:categories} and \ref{f}, \Tabref{tab:words})}. 

\minor{We discuss the relationship between three distributions: all foods consumed by the general population, food consumption estimated via MyFoodRepo, and food consumption estimated via Twitter.} The fact that there is a divergence between food consumption measured via the two platforms---food tracking and social media---implies that at least one of the two is not a faithful representation of true food consumption in the general Swiss population. \minor{We argue that it is less likely that food tracking is the main source of bias, and we conclude that researchers should be attentive and try to establish evidence of validity before using {digital traces} as a proxy for the true food consumption in the general population.} 

Measuring biases in digital traces is the first step towards correcting them and drawing valid conclusions despite their presence~\cite{wang2015forecasting}. \rev{Through a case study of the Twitter and MyFoodRepo platforms in Switzerland, contrasting tweeted food images with consumed and tracked foods, we provide grounding and first insights by controlling for location, period, and food types. Our findings cannot, however, be assumed to generalize globally, and future work should apply our framework to other populations, other social media platforms and Web traces, and other food tracking apps. Our study may serve the purpose of a ``proof by counterexample'': we have identified one common setting where there is a bias between two types of digital trace data. Hence, we should assume that there can be bias in other populations and platforms, too.}

We conclude the paper with a discussion (\Secref{sec:disscusion}) of how the methods and findings reported here can inform researchers in their efforts to leverage digital traces \rev{for various applications, in the context of food and beyond.}

\section{Background and Related Work}
\label{sec:related}

\subsection{Estimating food consumption from digital traces} 
We start by reviewing related work leveraging social media to study nutrition and dietary behaviors. Studying diets through social media posts has been an active area of CSCW research. Instagram \cite{phan2017healthy,garimella2016social,sharma2015measuring,ofli2017saki} 
and Twitter \cite{mejova2015twitter,abbar2015you,mejova2016fetishizing,mejova2015dietary,widener2014using,de2016characterizing} have emerged as particularly promising platforms.
Researchers have studied specific dietary issues and harmful behaviors. In particular, in work with important implications for the health and well\hyp being of vulnerable populations, researchers studied 
reports of eating disorders \cite{hunger2016,pro_eating2016}, 
dietary choices, nutritional challenges in food deserts (places with poor access to healthy and affordable food) \cite{de2016characterizing}, 
and obesity patterns in online behaviors \cite{mejova2015foodporn}. Related work has also studied eating disorder support online communities, 
quantifying and predicting disease severity and recovery \cite{de2015anorexia,chancellor2016thyghgapp}. 

Although social media has emerged as a rich data source, food shared or discussed on Instagram and Twitter might not be representative of food that people actually consume. Researchers have compared, at a population scale, statistics extracted from tweet text with public health statistics regarding the prevalence of obesity and diabetes \cite{abbar2015you,sajadmanesh2017kissing,mejova2015foodporn}. However, the content of posted images and the foods themselves have not been contrasted with actually consumed foods to date.

Beyond social media, researchers have long been applying their expertise to analyze health and nutrition behaviors using other kinds of digital trace data. First, researchers monitor food consumption with smartphone tracking applications and wearables \cite{achananuparp2016extracting,info:doi/10.2196/20625}. \minor{
Researchers analyzed compliance and contextualization of such platforms by investigating perceived and true snacking and meal consumption \cite{biel2018bites} and the potential for inferring population\hyp level eating routines \cite{gatica2019discovering}.}

Second, more distant proxies were previously used to analyze nutrition behaviors, including search engine logs \cite{West:2013:CCI:2488388.2488510,vosen2011forecasting,gligoric2022population},
purchase logs \cite{aiello2020tesco,aiello2019large,buckeridge2014method,gligoric2021ties,iwata2009topic,kawamae2010serendipitous,dijkstra2022cross},
online recipes \cite{wagner2014nature,wagner2014spatial,sajadmanesh2017kissing,rokicki2018impact,trattner2018predictability,tobey2019low},
reviewing platforms and websites \cite{chorley2016pub,weber2006internet,harris2012us,danescu2013no,ribeiro2021sudden}, 
crowdsourcing platforms \cite{Howell:2016:ATP:2896338.2896358,dunford2014foodswitch}, and geolocation signals \cite{sadilek2018machine}.
While food shared on social media might not be representative of consumed food, the above-listed, more distant proxies make it even harder to determine validity. For example, do recipe searches on search engines correspond to eating the food? Does reading an online recipe imply that the food was prepared and consumed? It is unknown to what extent such proxies imply food consumption, and it is not clear whether studies of food consumption via such digital traces truly measure the quantities intended to be measured.

\rev{In more distantly related research, researchers have been utilizing user\hyp generated food content to train and develop machine learning models. Current AI applications that use online food images include mining food photos to perform segmentation \cite{okamoto2021uec}, recognize food \cite{amato2017social,yanai2014twitter,sahoo2019foodai,bossard2014food}, learn food and recipe embeddings \cite{salvador2017learning}, and perform calorie \cite{naritomi2020caloriecaptorglass} and nutrient \cite{freitas2020myfood} estimation. However, if the food that people consume is systematically different from food shared online, models trained and evaluated on online datasets might not generalize to real\hyp world scenarios.}

\subsection{Biases of studying digital traces}

Next, we review related work studying biases of digital traces. The goal of measurements using behavioral trace data is to extract meaning from raw data that most often was not collected with the extraction of scientific insight in mind. Data\hyp driven research has thus been criticized for asking questions that appear to be opportunistically answerable with the data at hand, overlooking different types of biases \cite{gudivada2015big}.

\citeauthor{lazer2021meaningful}\ \cite{lazer2021meaningful} argue that the digital traces need to be linked to known constructs before we can use the data to answer scientific questions. Thus, the key challenge of studying digital data is determining whether measurements accurately capture the construct that one would ideally want to examine. For example, if one is measuring physical activity based on mobile phone location traces, how consequential is the omission of stationary activities such as treadmill or yoga \cite{lazer2021meaningful}? If one is tracking influenza with Web search logs of symptoms, how consequential are searches from persons not experiencing any symptoms \cite{lazer2014parable}?

The mismatch between the theoretical understanding of a concept and its operationalization, known as the issue of construct validity, can have harmful consequences \cite{wagner2021measuring}. In particular, 
when data that allows for measurement (\eg, arrest records) 
does not properly match the actual social construct that the measurement is intended to capture (\eg, a criminal act), 
measurements can replicate, mask, or exacerbate existing social issues \cite{corbett2017algorithmic}.

Related work has thus aimed to establish the validity of studying human behaviors with Web and social media traces. Example studies include studying the validity of screening depression \cite{kim2021systematic}, location traces \cite{horn2021investigating}, inferring political approval \cite{sen2020reliability}, sentiment analysis \cite{pellert2022validating}, or using Twitter's APIs \cite{morstatter2014biased}. \citeauthor{de2014seeking} \cite{de2014seeking} have studied seeking and sharing health information online by comparing search engines and social media. Researches have also studied decisions around whether to post content online \cite{10.1145/3209542.3209575}, political, racial and gender biases in Web systems \cite{kulshrestha2019search,hannak2017bias},
and how Web systems influence offline user behavior \cite{althoff2017online}. 

Further related work includes studies that issue calls to carefully scrutinize the use of social media data against biases and provide practical advice to aid researchers in performing their data\hyp driven studies. \citeauthor{sen2021total} \cite{sen2021total} proposed a total error framework for digital traces of human behavior on online platforms, \citeauthor{olteanu2019social} \cite{olteanu2019social} identified a variety of challenges in the practices of social media use for research, and \citeauthor{hofman2021integrating} \cite{hofman2021integrating} advocated for measuring the extent to which causal estimates made in one domain transfer to another domain. 
Whereas related work \cite{sen2021total,olteanu2019social,ruths2014social} aims to put the biases into a unified framework cutting through different domains, we aim to specifically establish the validity of estimating food consumption from \rev{digital traces}.

\section{Data and Methods}
\label{sec:methods}

\subsection{\rev{Food tracked via MyFoodRepo}}

To get as close as possible to capturing true food consumption, we use a novel dataset of food images collected via the MyFoodRepo mobile app \cite{mohanty2021food}. By design, the food present in these images was actually consumed, for the purpose of the application is to track users' personal food consumption. Through the app, volunteer users from Switzerland are asked to provide images of their complete daily food intake, mainly in the context of being enrolled in a digital cohort called Food \& You \cite{el2022associations}.\footnote{\url{https://www.digitalepidemiologylab.org/projects/food-and-you}} MyFoodRepo thus captures all foods that compliant individuals consume, in any context.

The images are publicly available as part of the Food Recognition Challenge.\footnote{\url{https://www.aicrowd.com/challenges/food-recognition-challenge}} The dataset has been annotated such that the individual foods are mapped onto an ontology of food types. Images were logged between 2017 and 2020. In our analyses, we study the training-set portion of the dataset, comprising 24,120 images, along with their corresponding 39,328 food-type annotations.

\subsection{\rev{Food shared on Twitter}}

To answer the question of whether images shared on social media diverge from food consumption \rev{as measured via food tracking}, we aim to contrast images of consumed \rev{and tracked} food with images of food posted on social media. To this end, we curate a dataset of food images shared on Twitter in Switzerland during the same period spanned by the images collected via the food tracking app, this way controlling for location and time.

Since our goal is to investigate the validity of studying diets with social media, in our Twitter data collection strategy, we, first, aim to follow data collection methods present in the existing literature closely, to be able to make conclusions that can be relevant for researchers working in this area, as opposed to inventing novel strategies that would be less relevant. Our data collection pipeline is therefore similar to pipelines described in related work, extracting nutritional information from social media posts with keywords (\Secref{sec:related}). Note that, since we follow existing work, specific data collection decisions are not limitations per se. Instead, the impact of data collection based on user-specified keywords is intended to be measured, since this is how researchers usually collect Twitter posts to study food consumption.

Second, we aim to gather a complete dataset, \ie, to collect all food images posted on Twitter by the relevant population in the relevant time frame. To this end, we use the full-archive search endpoint, available to researchers via Twitter's Academic Research product track,%
\footnote{\url{https://developer.twitter.com/en/docs/twitter-api/tweets/search/}} which allows searching Twitter's complete archive going back until March 2006.

Third, we aim to find images posted on Twitter that actually contain food, as we are interested in studying the posted food itself, rather than only how it is described. To this end, we apply automated and manual annotation of the collected images. With these three goals in mind, we employ the following data collection pipeline.

\xhdrIt{Step 1: Twitter data collection.}
We start from the set of \rev{MyFoodRepo} image annotations (\eg, ``bread'', ``banana''). We remove drinks and merge small types that are similar, obtaining 155 food types. We map each type to suitable handcrafted high-precision keywords, translate the keywords from English to German, French, and Italian (the large Swiss national languages) via Google Translate, and use the disjunction (``OR'') of keywords (separately per language) to query Twitter's full archive search API for the respective food type. We thus obtain all posts that in the text contain at least one of the keywords related to the food, in one of the four languages, in either singular or plural form (if relevant). For example, for the type ``bread'', we retrieve all English tweets containing ``bread'' or ``breads'', French tweets containing ``pain'' or ``pains'',
Italian tweets containing ``pane'' or ``pani'', and German tweets containing ``Brot'' or ``Brote''. Additional restrictions ensure that tweets were posted between 2017 and 2020 (the period when images of \rev{MyFoodRepo} food were logged) from a location in Switzerland and contain at least one image. This step yields 33,425 unique images.

\xhdrIt{Step 2: Automated annotation.} We are interested in studying the foods themselves, so we make sure that images indeed contain food. To that end, we perform detection of food in images with the ResNet50 model trained on ImageNet \cite{he2016deep}, which we finetuned for food-\vs{}-not-food classification on the publicly available Food-5K food image dataset \cite{singla2016food}, with 98\% recall and 96\% precision on the task of detecting food images on the held out 20\% test set (using a threshold of $p=0.5$).  Inspection of the images revealed that the images that do not contain food most frequently occur in food types where keywords have homonymous meanings. Two food types with the largest fraction of images that do not contain food are ``date'' (which can signify a fruit or ``day of a year'' or ``social appointment'') and ``apple'' (which can signify a fruit or Apple Inc.\ and its products). After this step, we keep 7,723 tweets with images that contain food.

\xhdrIt{Step 3: Manual annotation.} We manually inspect the images to verify that an image contains the food that the user mentions in the tweet text, even if a small quantity. \rev{The visible food item needs to be edible, \eg, a silver pendant of lemon shape or a carved and decorated Halloween pumpkin does not qualify. Additionally, the image needs to contain a prepared dish, and not all the ingredients laid out separately, nor an uncooked caught fish. Finally, no explicit content can be present in the background for the image to be safe for crowd workers.}

Due to the completeness of Twitter's full archive search and the manual inspection of collected images, at the end of the above process, we obtain all tweets posted from Switzerland between 2017 and 2020 with images that contain a food that is mentioned in the tweet text,
for a total of 3,692 images of food along with their corresponding 4,481 food-type annotations.

\begin{figure}[t]
\includegraphics[width =\columnwidth]{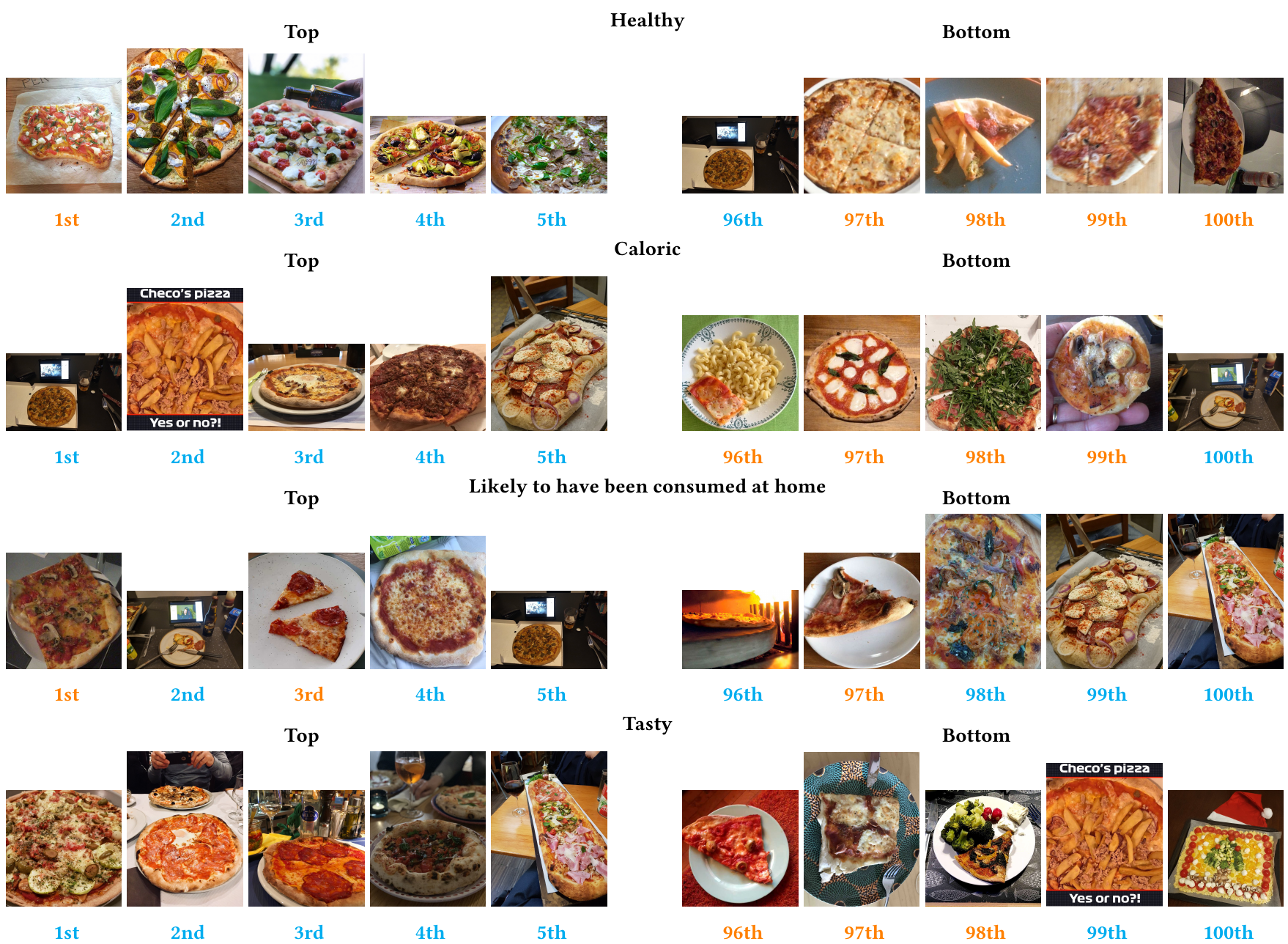}
\caption{\textbf{Example food images of type ``pizza''.} The overall top five (rank 1st--5th, \textit{left}) and overall bottom five (rank 96th--100th, \textit{right}) images with respect to estimated rank according to four criteria (one criterion per row). \rev{Twitter} foods are marked in blue and \rev{MyFoodRepo} foods, in orange.}
\label{fig:exampleimages}
\end{figure}

In summary, the two datasets we analyze contain $24{,}120$ images of \emph{consumed \rev{and tracked} food} and $3{,}692$ images of \emph{tweeted food}. Images are mapped on the food-type level and contain foods that we can compare in order to address our research questions. See \Figref{fig:exampleimages} for examples of images of type ``pizza''.

Having described the data, we continue by outlining our crowdsourcing framework for measuring biases. We then describe how we implement this framework on Amazon Mechanical Turk.

\subsection{Crowdsourcing framework for estimating biases}
Beyond food types, previous work (\Secref{sec:related}) has most notably used social media to estimate nutritional properties of food \cite{abbar2015you}, as well as its appearance and perception \cite{mejova2015foodporn}. Based on these themes, we operationalize four pertinent dimensions along which we contrast tweeted and consumed \rev{and tracked} food, capturing how
(1)~healthy,
(2)~tasty,
(3)~caloric, and
(4)~likely to have been consumed at home
the food is.

For each dimension, we aim to estimate a score for each image. Contrasting the scores of tweeted \vs\ consumed \rev{and tracked} food then allows us to assess biases.
In principle, we could obtain scores directly via human annotation by asking, \eg, ``How tasty does this dish look, on a scale from 1 to 10?'' It is, however, challenging for humans to place items on an interval scale that is consistent across individuals \cite{agresti2003categorical}. Based on the fact that judging between two alternatives is generally easier and more intuitive for humans \cite{chen2013pairwise}, we instead adopt a pairwise paradigm, where we confront human raters with pairwise choices (\eg, ``Which of these two dishes looks tastier?'') and later infer latent scores from the pairwise preferences.

Consider a given dimension (we use tastiness for concreteness in the following exposition) and a given food type.
Then, for two images $a$ and $b$ showing food of the same type, we use the notation ``$a \succ b$'' to express that $a$ is preferred over $b$ by a human rater.
Note that human preference is a random variable: different raters may have different preferences with respect to a given pair. We assume, however, that certain images show inherently tastier dishes and are thus more likely to be preferred.
More formally, following the Bradley--Terry (BT) model \cite{bradley1952rank}, we assume that each image $i$ has a latent tastiness score $s(i)$ and that the probability that a rater will prefer image $a$ over image $b$ [image $b$ over image $a$] in a pairwise comparison is proportional to the score of $a$ [score of $b$]:
\begin{equation}
        \Pr(a \succ b) = \frac{s(a)}{s(a) + s(b)} \label{eq:1}. 
\end{equation}
Given this setup, maximum likelihood estimation \cite{maystre2015fast} can be used in order to infer the latent scores that best explain the empirically observed pairwise preferences.
Thus, although only pairwise choices are made by humans, we can rank all images in a total order based on their latent scores $s$.
The BT model is appropriate for our purposes, as it has a well\hyp understood interpretation and is well\hyp suited to model human preferences \cite{chen2013pairwise}.
In practice, we fit a so-called Plackett--Luce model \cite{maystre2015fast}, a generalization of BT that does not require comparisons for all image pairs.



\subsection{Implementation on Amazon Mechanical Turk} In the remainder of this section, we describe how we implemented the above\hyp described framework on Amazon Mechanical Turk. 
To estimate the latent scores of images of \rev{MyFoodRepo} and of \rev{Twitter} food in terms of the four dimensions, we selected 24 well\hyp represented food types (\Figref{fig:medians}). The types are selected such that each type has at least 50 tweeted and at least 50 consumed \rev{and tracked} food images. The different food types are considered as independent ``tournaments'', so we obtained a separate ranking per type.

We first sampled the same number of tweeted food images and consumed \rev{and tracked} food images per type, to account for potentially different food-type frequencies. Recall that location and period are already controlled for in the data collection. We randomly sampled 100 images from each type, 50 tweeted images and 50 images of consumed \rev{and tracked} foods, resulting in 2,400 competing images in total.

\begin{figure}[t]
        \includegraphics[width=0.5\textwidth]{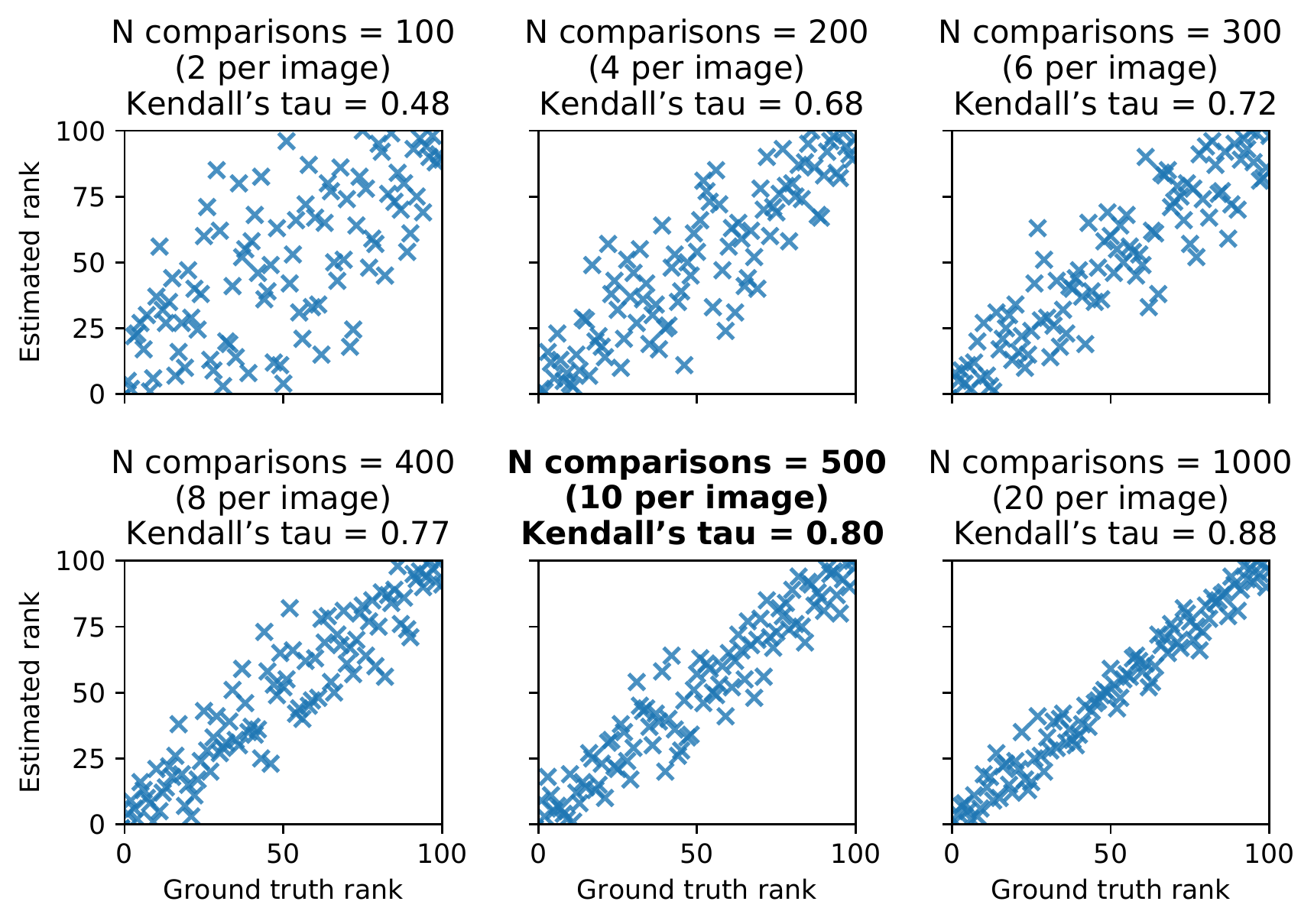}
    \caption{\textbf{Determining a sufficient number of comparison pairs via rank reconstruction simulations.} Ground truth rank ($x$\hyp axis) and estimated rank ($y$\hyp axis), for varying number of comparisons ($N$). In our subsequent analysis, we chose to perform 10 comparisons per image ($N=500$).}
    \label{fig:simulations}
\end{figure}

We then performed random sampling of comparison pairs. From each set of images for a given food type, we sampled $N$ pairwise comparisons, each time randomly sampling one image of consumed \rev{and tracked} food and one image of tweeted food, constrained such that each image participates in the same number of comparisons. We chose the number of ``duels'' per food type based on rank inference simulations, with the goal of ensuring that we can infer true ranks accurately, as follows.
\rev{We assumed 100 items divided into two groups with item quality sampled from the standard normal distribution. We ranked the items and then randomly sampled $N$ comparisons between items from the two groups. We sampled the outcome of a duel based on the items' quality scores and estimated quality and rank with a BT model (\Eqnref{eq:1}). \Figref{fig:simulations} depicts how well the true ranking can be recovered for different numbers $N$ of comparisons. As more comparisons are performed, the estimated rank ($y$-axis) correlates more strongly with the ground-truth rank ($x$-axis). Based on these results, we chose to perform 10 comparisons per image ($N=500$, Kendall's $\tau=0.80$). That is, at $N=500$, each of the 50 images is compared to 10 competitors, and rank can be accurately inferred with Kendall's $\tau=0.80$.}

In every rating task, a participant was shown a random pair of images containing food of the same type (images in a pair were scaled to the same size and shown in randomized order), and asked to give a preference label for each of the four dimensions (healthiness, tastiness, caloric content, likelihood to be consumed at home). The pairwise comparison task had no neutral option; participants were required to choose one image. As the order within pairs was randomized, this is a valid way of breaking ties, and recommended practice \cite{perez2017practical}. Additionally, we asked participants to explain how they perceived both images by providing between one and three free-form tags (\eg, ``dull'', ``greasy'').
(Prior to data collection, we did not make hypotheses about specific biases as revealed by the tags, but rather explore them post\hyp hoc in order to gain insights about how people describe the appearance of tweeted \vs\ consumed \rev{and tracked} food.) In total, we collected 12,000 pairwise comparisons (500 duels for each of 24 types) for each of the four dimensions.

\subsubsection{Participants} Since the tasks require reading and writing text in English, participants were restricted to those residing in the United States, Canada, or the United Kingdom. To ensure high-quality answers, we admitted only workers with approval rates greater than 99\% and with more than 1,000 previously approved tasks. We collected the 12,000 pairwise preferences through 24 batches with 500 assignments each, over the course of five days. The task was performed by 595 distinct workers, who performed 20.2 pairwise comparisons each, on average.

\subsubsection{Compensation} We targeted a pay rate of \$9 per hour. Participants were paid \$0.15 per pairwise comparison. The mode of the time taken per comparison was 57 seconds, which corresponds to an estimated hourly rate of \$9.50 (the U.S.\ federal minimum hourly wage in 2021 was \$7.25  per hour, for reference). Note that this is likely an underestimate of the hourly rate since crowd workers often use scripts that make it possible to automatically accept a task they are interested in, and hold it assigned while not actively working on it. 

\subsubsection{Instructions} To ensure reproducibility of our experiment, below we quote the instructions as they were displayed to participants:

\vspace{5mm}
\noindent \begin{center} \fbox{%
    \small
    \parbox{0.85\linewidth}{\raggedright

     \textit{Please take a look at the two images displayed below. Please focus on the food itself, and not the other contents of the image.} 
\textit{Answer the questions about the pizza shown in the images by entering either \textbf{1} for Image 1, or \textbf{2} for Image 2. Additionally, please explain your preferences by adding \textbf{at least one word or short phrase to describe the foods} appearing in each image. Write a word or a short phrase, and not full sentences. }\\

\vspace{0.5cm}
1: Which image contains pizza that appears more \textbf{tasty}? 

2: Which image contains pizza that appears more \textbf{healthy}? 

3: Which image contains pizza that appears more \textbf{caloric}?

4: Which image contains pizza that appears more likely to have been \textbf{consumed at home}?

5: Pizza shown in Image 1 is \dots\\
 (Add a word or a short phrase to describe food in Image 1)

6: Pizza shown in Image 2 is \dots\\
 (Add a word or a short phrase to describe food in Image 2)
    }%
}\end{center}


\section{Results}
\label{sec:results}
\subsection{RQ1: Bias of food-type distribution}
\label{sec:RQ1}

To address RQ1, we start by comparing \rev{MyFoodRepo} food images with images of foods posted on Twitter. We compare the prevalence of the 155 food types across the two sets (\Figref{fig:categories}).

First, we observe a significant positive correlation (Spearman's rank correlation coefficient $\rho=0.49, p=8.5\times 10^{-11}$). The more frequent a food type is among consumed \rev{and tracked} foods, the more frequent it tends to be among tweeted foods. That said, although food type frequencies are correlated, important deviations can be observed.
For instance, bread and butter are more likely to be observed in \rev{MyFoodRepo} foods compared to \rev{Twitter} foods, \ie, these foods are underrepresented on Twitter. Bread is 2.5 times more frequent among \rev{MyFoodRepo} foods, while butter is 5.5 times more frequent among \rev{MyFoodRepo} foods. On the other hand, cake, soup, chocolate, raclette, burgers, \etc, are more likely to be observed among \rev{Twitter} foods, compared to \rev{MyFoodRepo} foods, \ie, these foods are overrepresented on Twitter. Soup is 11.5 times, cake 12.0 times, burger 10.0 times, and raclette 9.5 times more frequent among \rev{Twitter} foods.


\subsection{RQ2: Biases within food types}
\label{sec:RQ2}

\subsubsection{\rev{Duel outcomes}} 
\begin{figure*}
    \begin{minipage}[b]{0.55\textwidth}
    \centering
    \hspace{-1.7cm}
    \includegraphics[width=\textwidth]{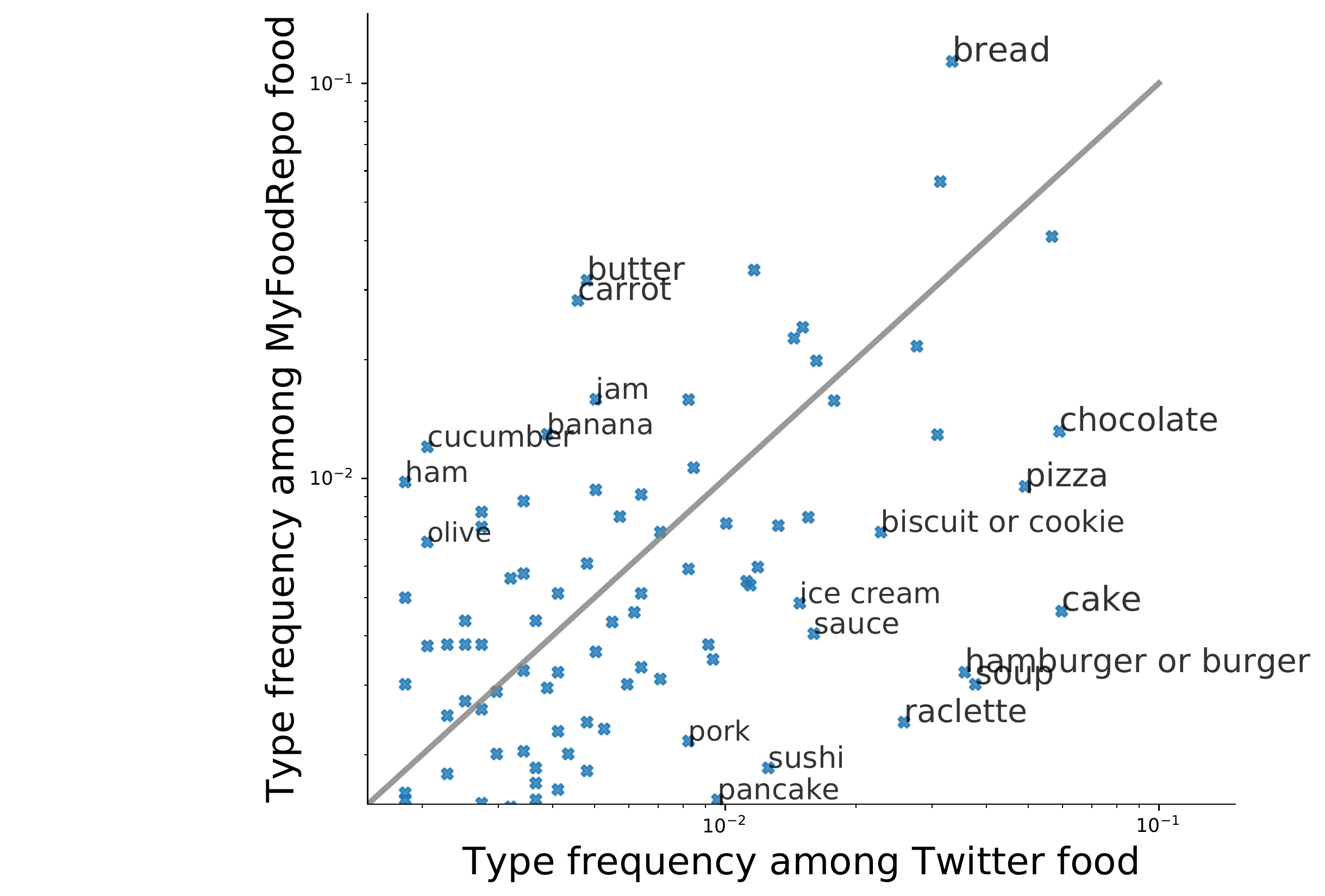}
    \subcaption{}
    \label{fig:categories}
    \end{minipage}
    \hfill
    \begin{minipage}[b]{0.44\textwidth}
    \centering

    \includegraphics[width = \textwidth]{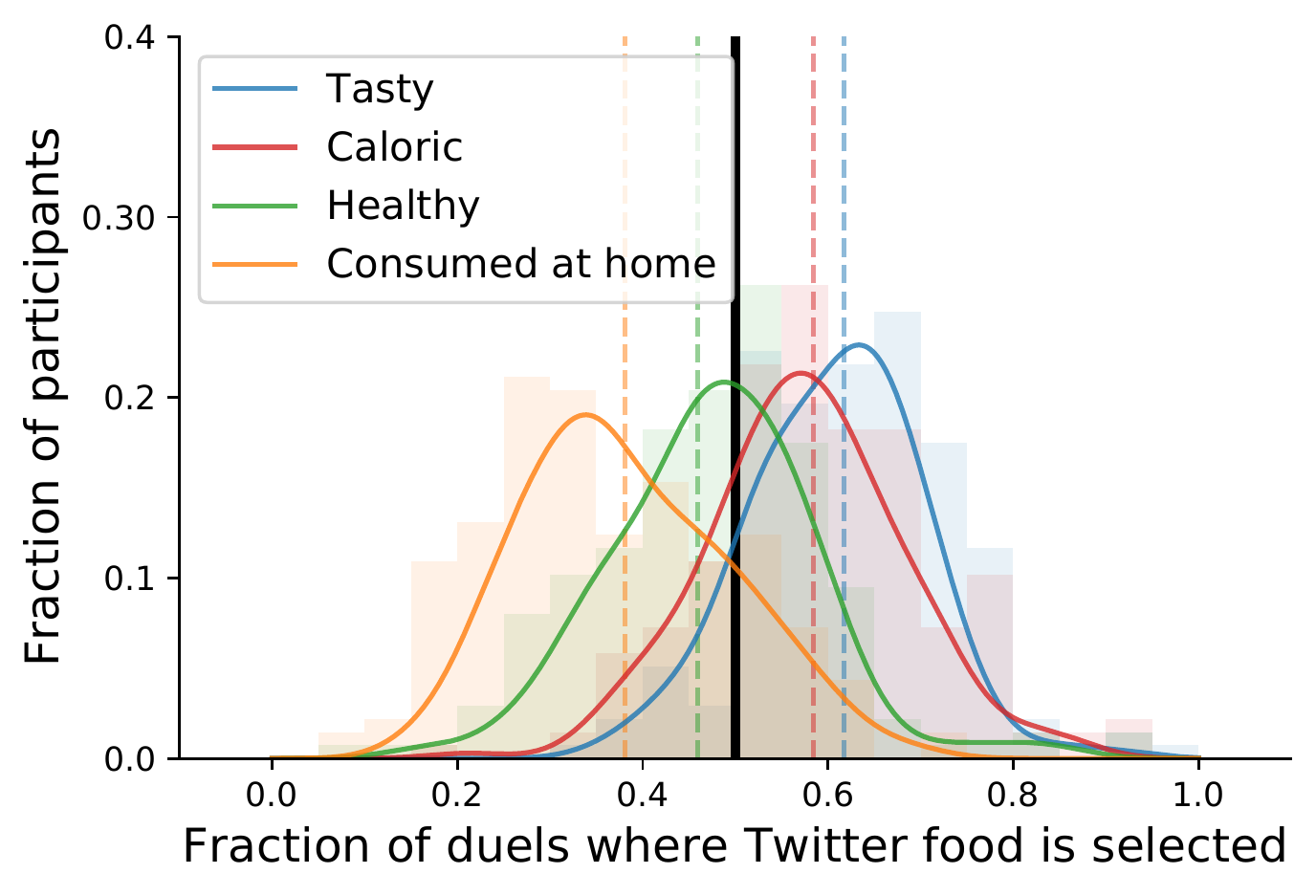}
    \subcaption{}
    \label{fig:dist_workers}
    \end{minipage}
\caption{\textbf{Bias of food-type distribution, and within food-type preferences.} (a) Comparison of food type frequency among \rev{Twitter} food ($x$\hyp axis) \vs\ \rev{MyFoodRepo} food ($y$\hyp axis). \rev{Categories where the larger frequency is at least two times greater than the smaller frequency are annotated.}
Gray diagonal line marks identity, where two frequencies are equal. (b) Histogram of crowd workers' preferences. On $x$-axis, fraction of duels where \rev{Twitter} food image is selected when compared to a \rev{MyFoodRepo} food image. On $y$-axis, fraction of workers with such preferences. Dashed vertical lines mark average fractions of wins, across participants. Solid vertical line marks $0.5$. 
}
\label{}
\end{figure*}

Controlling for the different food-type distributions, we address RQ2, which is concerned with biases within fixed food types. \rev{As an initial look into the duel outcomes, across all duels, we first consider the fraction where the Twitter image won. Together with this fraction, we report $p$\hyp values from two\hyp sided binomial tests, where the null hypothesis is that the outcome of comparisons is random, \ie, that the Twitter image wins in 50\% of duels. 
Across all duels, 
in 58.46\% of duels the Twitter image is chosen as more caloric $(p <10^{-70})$, 
in 45.96\% as more healthy $(p <10^{-10})$, 
in 38.08\% as more likely to have been consumed at home $(p<10^{-140})$, 
and 
in 61.73\% as more tasty $(p <10^{-100})$.}%
\footnote{We also examined the macro\hyp average duel outcomes, where we consider the preference of each crowd worker and then average over workers. There appears to be no rater bias where some workers overwhelmingly prefer one or the other, as the estimates are consistent, and no notable outlier crowd workers emerge (\Figref{fig:dist_workers}).}

\subsubsection{\rev{Bias measurement: score estimations}} \rev{Next, in order to compare images, as opposed to the outcomes of duels, we fit the BT model (\Eqnref{eq:1}) on the collected preferences and estimate a score that represents the latent quality of each competing image concerning how healthy, tasty, caloric, and likely to have been consumed at home the food appears.} 

\rev{Consider a given dimension (we again use tastiness for concreteness). Let each MyFoodRepo image $i\in \{ 1,...,N_M \}$ have a an estimated tastiness score $\hat s(i)$,
and each Twitter image $j\in \{1,...,N_T \}$ have an estimated tastiness score $\hat s(j)$.
The tastiness bias $b(T,M)$ between food consumption measured with Twitter and food consumption measured with MyFoodRepo can be expressed as the difference in the average estimated tastiness scores measured via the respective data sources, $T$ and $M$:}

\begin{equation}
        \rev{b(T,M)  = T - M = {\frac {1}{N_T}}\sum _{i=1}^{N_T}\hat s(i) - {\frac {1}{N_M}}\sum _{j=1}^{N_M}\hat s(j)  \label{eq:2}.}
\end{equation}

\rev{The measured bias $b(T,M)$ (with 95\% confidence intervals obtained via bootstrap resampling of images) is
$0.52$ $[0.46, 0.56]$ for ``tasty'',
$0.39$ $[0.35, 0.45]$ for ``caloric'',
$-0.18$ $[-0.23, -0.11 ]$ for ``healthy'', and
$-0.58$ $[-0.63, -0.52 ]$ for ``likely to have been consumed at home''. These bias measurements indicate that, on average, food posted on Twitter is
perceived as significantly tastier, more caloric, less healthy, and less likely to have been consumed at home, compared to consumed and tracked food.} 
As estimated scores are not located on an interpretable scale, we shift our focus to ranks instead of raw scores next.

\subsubsection{\rev{Bias measurement: rank estimations}} \rev{Once latent scores are estimated, images can also be ranked, either jointly, or separately, among Twitter and \rev{MyFoodRepo} food. Given its intuitive interpretation,} our main method of analysis is quantifying the shifts in distributions via ranks, as depicted in \Figref{fig:diagram}. For concreteness, it is helpful to consider an example before studying images on a more aggregate level across food types. \Figref{fig:exampleimages} contains the top and bottom portions of the joint rankings (one ranking for each of the four dimensions) for food type ``pizza''.

\begin{figure*}
    \begin{minipage}[t]{0.81\textwidth}
    \centering
    \includegraphics[width=\textwidth]{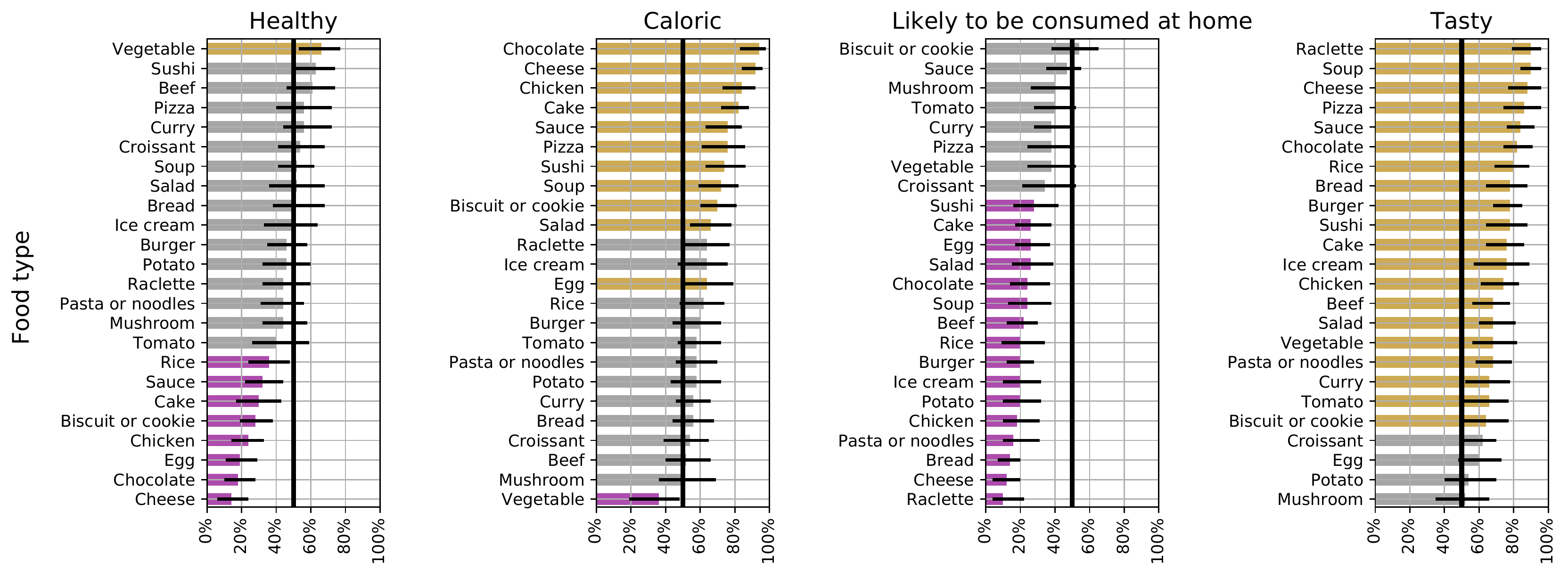}
    \subcaption{}
    \label{fig:medians}
    \end{minipage}
    \hfill
    \begin{minipage}[t]{.18\textwidth}
    \centering
    \includegraphics[width = \textwidth]{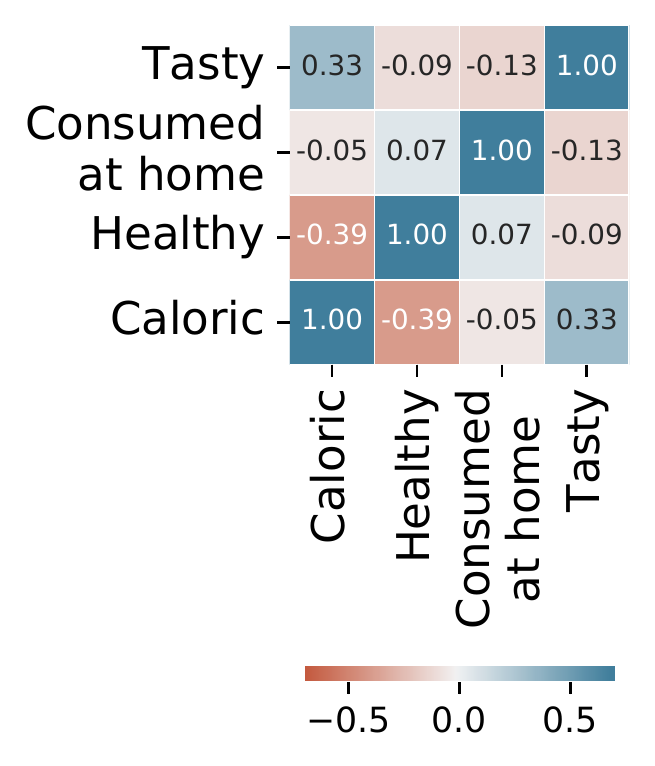}
    \subcaption{}
    \label{fig:corr}
    \end{minipage}
\caption{\textbf{Rank bias estimations and correlation between estimated qualities.}
(a) Relative rank (with respect to estimated latent scores) of median tweeted food among consumed \rev{and tracked} foods, where 100\% corresponds to top score. Colored bars mark estimates that significantly differ from median (\ie, 50\% relative rank). Yellow marks ranks higher, purple marks ranks lower than median, while gray marks non\hyp significant differences. Error bars mark 95\% confidence intervals obtained via bootstrap resampling of duels. Example: relative rank of median tweeted raclette with respect to tastiness is 90\% (95\% CI $[79\%,96\%]$); \ie, median\hyp tasty tweeted raclette is ranked among top 10\% $[4\%,21\%]$ tastiest consumed \rev{and tracked} raclette images. (b)~Correlation matrix between estimated qualities among 2,400 competing images.
\unboldmath
}
\label{f}
\end{figure*}

Now, for each food type, we rank the two sets (\rev{MyFoodRepo} food \vs \rev{Twitter} food) separately and determine to which percentile of \rev{MyFoodRepo} food each percentile of \rev{Twitter} food corresponds. We focus on the median \rev{Twitter} food image (also referred to as the ``typical'' \rev{Twitter} image), and compute the percentile rank of its score, relative to scores of \rev{MyFoodRepo} food images. The rank of the median \rev{Twitter} image among \rev{MyFoodRepo} food images is presented in \Figref{fig:medians}, across the 24 food types.
             
\xhdrIt{Bias in nutritional properties}
We find that \rev{Twitter} foods are perceived as more caloric, less healthy, and less likely to have been consumed at home. On average across food types, 
the median-caloric \rev{Twitter} food is among the top 34\% most caloric \rev{MyFoodRepo} foods. 
The median-healthy \rev{Twitter} food is among the bottom 42\% most healthy \rev{MyFoodRepo} foods. 
And finally, the median \rev{Twitter} food is among the bottom 27\% most likely home-consumed \rev{MyFoodRepo} foods.

Examining food types separately, regarding how healthy the foods are estimated to be, we find no significant differences in 15 out of the 24 types. In eight food types, tweeted food is perceived as \emph{less healthy}, and one food type (vegetables) is found to be more healthy on Twitter. With respect to perceived caloric content, we find no significant differences in 12 out of the 24 types. For 11 types, tweeted food is perceived as {more caloric}. Vegetables are again an exception, found to be less caloric on Twitter compared to \rev{MyFoodRepo} foods. We find that in most of the types (16 out of the 24), tweeted foods are {less likely to have been consumed at home}. For eight food types, there are no significant differences in likelihood of having been consumed at home.

Although there are food types with no biases in nutritional properties, we observe large biases for certain foods, where a median-healthy \rev{Twitter} food is among the bottom 20\% of \rev{MyFoodRepo} food images with respect to healthiness. For instance, median-healthy \rev{Twitter} cheese is among the bottom 14\% (95\% confidence interval $[6\%, 24\%]$) \rev{on MyFoodRepo};
median-healthy \rev{Twitter} chocolate, among the bottom 18\% $[10\%,28\%]$ \rev{on MyFoodRepo};
and median-healthy eggs, among the bottom 19\% $[11\%,29\%]$  \rev{on MyFoodRepo}.

Moreover, the median-caloric \rev{Twitter} food is among the top 20\% caloric \rev{MyFoodRepo} food. For example, median-caloric chocolate is among the top 6\% $[2\%,17\%]$ \rev{on MyFoodRepo};
median-caloric cheese, among the top 8\% $[4\%,16\%]$ \rev{on MyFoodRepo}; 
median-caloric chicken, among the top 16\% $[8\%,27\%]$ \rev{on MyFoodRepo};
and median-caloric cake, among the top 18\% $[12\%,28\%]$ \rev{on MyFoodRepo}.

\xhdrIt{Bias in tastiness}
Next, we find substantial bias in how tasty the foods are perceived to be. There are more discrepancies in perceived tastiness compared to the above nutritional properties (\Figref{fig:medians}). 
On average across food types, the median-tasty tweeted food is among top 26\% tastiest consumed \rev{and tracked} foods. A median-tasty tweeted food is ranked significantly higher than the median-tasty consumed \rev{and tracked} food image in 20 out of the 24 types. In only four types there are no significant differences regarding tastiness (mushrooms, potato, egg, croissant).

We note that, for a number of foods, a median-tasty \rev{tweeted} food is ranked as high as among the top 20\% of consumed \rev{and tracked} food. The median-tasty \rev{Twitter} raclette is among the top 10\% $[4\%, 21\%]$ \rev{on MyFoodRepo};
the median-tasty soup, among top 10\% $[4\%, 16\%]$ \rev{on MyFoodRepo};
the median-tasty cheese, among top 12\% $[4\%, 23\%]$ \rev{on MyFoodRepo};
the median-tasty pizza, among top 14\% $[4\%, 26\%]$ \rev{on MyFoodRepo};
the median-tasty sauce, among top 16\% $[18\%, 24\%]$ \rev{on MyFoodRepo};
and the median-tasty \rev{Twitter} chocolate, among top 18\% $[9\%, 26\%]$ \rev{on MyFoodRepo}.

\subsubsection{Correlations} Next, we inspect the correlation between tastiness and the other nutritional properties. Recall that we obtained 2,400 images (100 images for each of the 24 types), with four estimated quality scores. Computing Pearson's correlation between the estimated scores (\Figref{fig:corr}), we observe a moderate positive correlation between how caloric and tasty ($\rho=0.33$, $p<10^{-60}$) foods are, and a negative correlation between how caloric and how healthy ($\rho = -0.39$, $p<10^{-80}$) they are.
Whereas the negative correlation between how caloric and how healthy foods are is expected, the correlation between how tasty and how caloric they are might indicate that tweeted food might be perceived as overwhelmingly tastier because it is more caloric and exaggerated.

\subsubsection{Complete rank comparisons} 
\begin{figure*}
    \centering
    \hspace{0.2cm}
    \includegraphics[width=\textwidth]{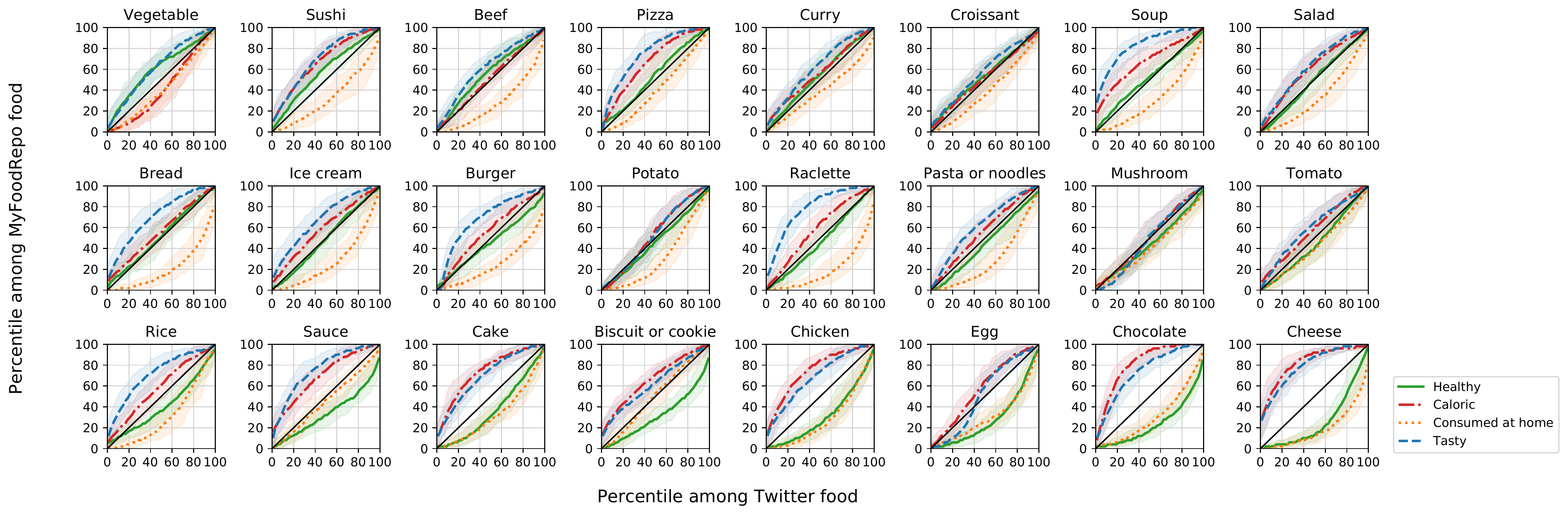}
    \caption{\textbf{Complete rank comparisons across food types.} For each percentile among \rev{Twitter} food (on the $x$-axis), the corresponding percentile among \rev{MyFoodRepo} food (on the $y$-axis). That is, a Twitter image at percentile rank $x$ among Twitter images would place at percentile rank $y$ if it were ranked against MyFoodRepo images instead. Error bars mark 95\% confidence intervals obtained via bootstrap resampling. Rank comparison is displayed for 24 types of food, across four estimated qualities: how healthy, caloric, likely to have been consumed at home, and tasty the food is. Food types are sorted according to the rank of the median\hyp healthy tweeted food among consumed \rev{and tracked} foods. Black diagonal line marks identity, where percentile distributions of \rev{Twitter} food and \rev{MyFoodRepo} food are equal.}
    \label{fig:cumulative}
\end{figure*}

\rev{Whereas so far we focused on the median Twitter image and studied its relative rank among \rev{MyFoodRepo} images, in \Figref{fig:cumulative} we present the full percentile rank comparisons, for completeness. The black diagonal line marks identity, where the percentile distributions of Twitter food and \rev{MyFoodRepo} food are equal. In the example of cheese (bottom right), percentiles among tweeted food correspond to higher percentiles among consumed \rev{and tracked} food regarding how caloric and tasty the cheese is (tweeted food ranks are above the diagonal line). Percentiles among tweeted food correspond to lower percentiles among consumed \rev{and tracked} food regarding how healthy and home-consumed the cheese is (tweeted food ranks are below the diagonal line).}

\subsubsection{Bias in appearance}
\label{sec:Bias in appearance}
Finally, we analyze the tags provided by crowd workers, in order to understand further how \rev{MyFoodRepo} food and \rev{Twitter} food differ in their appearance.%
\footnote{Note that this is an exploratory post\hyp hoc analysis to gain deeper insights into potential mechanisms that drive the observed biases. Our main analyses are related to perceived nutritional properties and tastiness, as estimated via pairwise comparisons.}
Recall that crowd workers were asked to enter up to three tags per image (on average, participants entered 2.4 tags per image). Analyzing the 58,645 collected tags, we ask: \emph{How do people perceive social media foods compared to consumed \rev{and tracked} foods?}

We first performed normalization of the tags provided by crowd workers. We split commas, convert to lowercase, remove stop\hyp words at the beginning of the text (\eg, ``looks'', ``seems''), and we map versions of words with a dash to a single form (\eg, mapping ``mouth watering'' and ``mouthwatering'' to ``mouth\hyp watering'').

A tag is typical for one set of images if used frequently within the set, but at the same time unlikely to be used in the other set. Additionally, it is not only the discrepancy between the two probabilities that matters; a tag should also appear frequently in a set to be considered typical of the set. This intuition is captured by the pointwise Kullback--Leibler (KL) divergence
between the distributions of tags for \rev{Twitter} food images and for \rev{MyFoodRepo} food images, respectively. Specifically, the distinctiveness of a tag $t$ with respect to \rev{MyFoodRepo} food compared to \rev{Twitter} food images is calculated as
\begin{equation}
       \KL(\pmfr(t)\|\pt(t)) = \pmfr(t) \log \frac{\pmfr(t)}{\pt(t)}, \label{eq:3} 
\end{equation}
where $\pmfr(t)$ is the probability of observing tag $t$ among \rev{MyFoodRepo} food images, and  $\pt(t)$ the probability of observing $t$ among \rev{Twitter} food images. On the other hand, since the KL divergence is not symmetric, the distinctiveness of \rev{Twitter} food compared to \rev{MyFoodRepo} food is calculated as
\begin{equation}
\begin{aligned} 
       \KL(\pt(t)\|\pmfr(t)) = \pt(t) \log \frac{\pt(t)}{\pmfr(t)}. \label{eq:4} 
\end{aligned}
\end{equation}

\begin{table}
\footnotesize

\caption{\textbf{Bias in appearance.} 
Tags most distinctive of \rev{MyFoodRepo} food (\textit{left})
or of \rev{Twitter} food (\textit{right}),
as determined by pointwise KL divergence (cf.\
\Secref{sec:Bias in appearance}%
);
*$p<0.05$, **$p<0.01$, ***$p<0.001$, ****$p<0.0001$).
\unboldmath
}
\label{tab:words}
\setlength\tabcolsep{3pt}
\begin{tabular}{l|lll|lll}
\toprule

\textbf{Food} & \multicolumn{3}{c}{\textbf{Top tags typical for \rev{MyFoodRepo} food}} \vline & \multicolumn{3}{c}{\textbf{Top tags typical for \rev{Twitter} food}} \\
\midrule
& plain**** &
bland**** & 
simple**** &
fancy**** &
fresh****  &
colorful**** \\ 

\textbf{Overall} & 
boring**** &
small**** &
thin**** &

delicious &
tasty*** &
raw**** 
\\

& dry**** &
healthy*** &
homemade****  &
gourmet**** &
flavorful**** &
large**** 

\\ 

\midrule

\textbf{Beef}
& grilled**
& bland**
& simple*
& raw***
& fancy*
& flavorful
\\ 

\textbf{Biscuit or cookie}

& shortbread**
& chocolate*
& plain*
& decorated***
& festive**
& chocolate chip**
\\

\textbf{Bread}

& white****
& plain**
& dry**
& delicious**
& fresh**
& fancy*
\\ 

\textbf{Burger}

& simple***
& small size*
& fast food*
& interesting**
& attractive*
& filling*
\\ 

\textbf{Cake}

& small****
& chocolate****
& chocolatey****
& decorated****
& fruity***
& fancy**
\\ 

\textbf{Cheese}

& boring***
& plain***
& cold***
& melted****
& warm****
& hot****
\\ 

\textbf{Chicken}

& plain****
& bland****
& dry****
& fried***
& tasty***
& spicy**
\\ 

\textbf{Chocolate}

& dark****
& bitter***
& broken***
& delicious***
& flavoured**
& fancy**
\\ 

\textbf{Croissant}

& delicious*
& buttery*
& stuffed*
& warm*
& fresh
& healthy
\\ 

\textbf{Curry}

& rice**
& white*
& homestyle*
& yummy*
& healthy*
& flavorful*
\\ 

\textbf{Egg}

& plain***
& simple***
& bland**
& decorated**
& raw**
& colorful**
\\ 

\textbf{Ice cream}

& chocolate***
& nice*
& simple*
& delicious
& decadent
& fancy
\\ 

\textbf{Mushroom}

& colorful
& simple
& pizza
& raw**
& creamy
& large
\\ 

\textbf{Pasta or noodles}

& plain**
& bland**
& simple**
& fancy***
& delicious**
& yellow*
\\

\textbf{Pizza}

& small****
& pepperoni**
& saucy*
& large***
& fresh**
& delicious*
\\ 

\textbf{Potato}

& boiled****
& peeled****
& plain*
& raw***
& unpeeled**
& whole*
\\ 

\textbf{Raclette}

& healthy***
& greasy***
& unappetizing*
& sliced**
& hot**
& tasty*
\\ 

\textbf{Rice}

& bland***
& plain**
& dry*
& flavorful**
& mixed*
& fried*
\\ 

\textbf{Salad}

& simple**
& plain**
& homemade**
& filling*
& great*
& fresh
\\ 

\textbf{Sauce}

& thin****
& watery***
& light****
& thick****
& creamy****
& red****
\\ 

\textbf{Soup}

& bland****
& simple****
& watery***
& hearty***
& colorful***
& delicious**
\\ 

\textbf{Sushi}

& boring***
& simple**
& plain**
& appetizing*
& variety*
& fresh
\\ 
\textbf{Vegetable}
& chopped*
& overcooked*
& mixed
& fresh**
& raw*
& delicate
\\
\textbf{Tomato}
& small*
& healthy
& salad
& mouth-watering*
& juicy*
& flavorful

\end{tabular}
\end{table}

In \Tabref{tab:words}, we present the tags with the largest pointwise KL divergence, separately for tags distinctive of consumed \rev{and tracked} food (left) and of tweeted food (right). For each tag, a $\chi^2$ test on the two frequencies is used to measure significance, under the null hypothesis that the two groups do not differ in frequency.
We now examine---first overall, then separately by type---how foods differ in their appearance.
We see that, overall, the tags most indicative of consumed \rev{and tracked} food are ``plain'', ``bland'', ``simple'', ``boring'', ``small'', ``thin'', ``dry'', ``healthy'', and ``homemade''. On the contrary, tweeted food is more likely to be described as ``fancy'', ``fresh'', ``colorful'', ``delicious'', ``tasty'', ``raw'', ``gourmet'', ``flavorful'', and ``large''. 
Zooming into specific food types, the exact differences for specific food type become apparent. For example, tweeted pizzas are more likely to be described as ``large'', ``fresh'', and ``delicious'', whereas consumed \rev{and tracked} pizzas are seen as ``small'', ``pepperoni'', or ``saucy''.

Inspecting the most discriminative tags overall and separately across types of food, we identified the following four prominent themes in tags that are discriminative of consumed \rev{and tracked} \vs\ tweeted food:

\begin{enumerate}
\item \emph{Complexity.} Consumed \rev{and tracked} food is described as simple and homemade (``plain'', ``bland'', ``simple'' and ``homemade''); tweeted food, as more elaborate (``fancy'', ``gourmet'').
\item \emph{Portion size.} Consumed \rev{and tracked} food comes in small portions (``small'', ``thin''), whereas tweeted food is exaggerated in portion size (``large''). Portion size differences are particularly evident for specific types of food, \eg, consumed \rev{and tracked} burgers are described as ``small size''; pizzas are ``small'' when consumed \rev{and tracked}, and ``large'' when tweeted.
\item \emph{Ways of preparing.} Tweeted food is perceived as ``raw'' and ``fresh'', while consumed \rev{and tracked} food is described as ``dry''. The differences are evident when it comes to specific foods. Consumed \rev{and tracked} beef is more likely to be ``grilled'', whereas tweeted beef is more likely ``raw''. Consumed \rev{and tracked} vegetables are ``chopped'' and ``overcooked'', whereas on Twitter, they are ``fresh'' and ``raw''. Rice and chicken are ``fried'' on Twitter, and ``dry'' when consumed \rev{and tracked}.
\item \emph{Presentation.} Tweeted food is visually appealing, whereas consumed \rev{and tracked} food is more likely to look repulsive. Tweeted food is said to be ``colorful'', ``delicious'', ``flavorful'', and ``tasty'', whereas consumed \rev{and tracked} food is usually less appealing, with ``watery'' soup, ``greasy'' and ``unappetizing'' raclette, and ``watery'' and ``thin'' sauce.
\end{enumerate}

\section{Discussion}
\label{sec:disscusion}
Our goal has been to determine the validity of estimating food consumption from digital traces: from social media posts \vs from images of consumed and tracked foods. To this end, we designed a crowdsourcing framework for measuring biases, contrasting tweeted food images with consumed \rev{and tracked} foods images, and deployed it for the case of Switzerland.

\subsection{Summary of main findings}

We find that social media does not provide a faithful representation of food types \rev{of consumed and tracked food}. Measuring biases in food-type distributions, we observe that cake, soup, chocolate, raclette, and burgers are among the most overrepresented foods on Twitter in Switzerland. Cake, soup, and burger are visually appealing food types suitable for sharing on social media, while chocolate and raclette are foods typical of Switzerland. Winter social and sport activities among residents might make them more likely to be shared online. On the other hand, bread and butter---among the most underrepresented on Twitter---are simple everyday foods that tend not to have a lot of potential to look particularly visually appealing.

Controlling for the discrepant food-type distributions, we find that tweeted foods are perceived as less healthy, more caloric, and less likely to have been consumed at home, compared to consumed \rev{and tracked} food of the same type (\Figref{fig:medians}). We also find substantial bias in perceived tastiness. A median-tasty tweeted food is, on average across food types, ranked among the top 26\% of consumed \rev{and tracked} foods. Exploring free-form tags provided by crowd workers reveals that these biases are likely mediated by differences in portion size, complexity, presentation, and different ways of preparing food. For example, tweeted food is 3.5 times more likely to be described as ``large'', and 4 times more likely to be described as ``fancy''.

These results provide evidence that food shared online tends to be exaggerated compared to tracked food. The most biased foods in terms of nutritional properties are foods that can be very caloric and high in fat and carbohydrates: chocolate, cheese, chicken, cake, and egg. On the other hand, we find that some of the foods are not skewed in terms of nutritional properties. For instance, for mushrooms and croissants, no significant difference is observed in any of the four dimensions (healthiness, tastiness, caloric content, likelihood to be consumed at home). This suggests that some foods can still be validly studied via social media \rev{as a proxy for consumed and tracked foods}. 

\subsection{\rev{Consumed food \vs tracked food \vs tweeted food}}

\rev{Our study attempts to establish a link between online and offline dietary behaviors by studying food images as measured via two platforms: Twitter and the MyFoodRepo food\hyp tracking app. In what follows, we consider the relationship between three distributions: all foods consumed by the general population (the actual phenomenon of interest), food consumption estimated via MyFoodRepo, and food consumption estimated via Twitter.}

\begin{figure*}
    \centering
    \includegraphics[width=\textwidth]{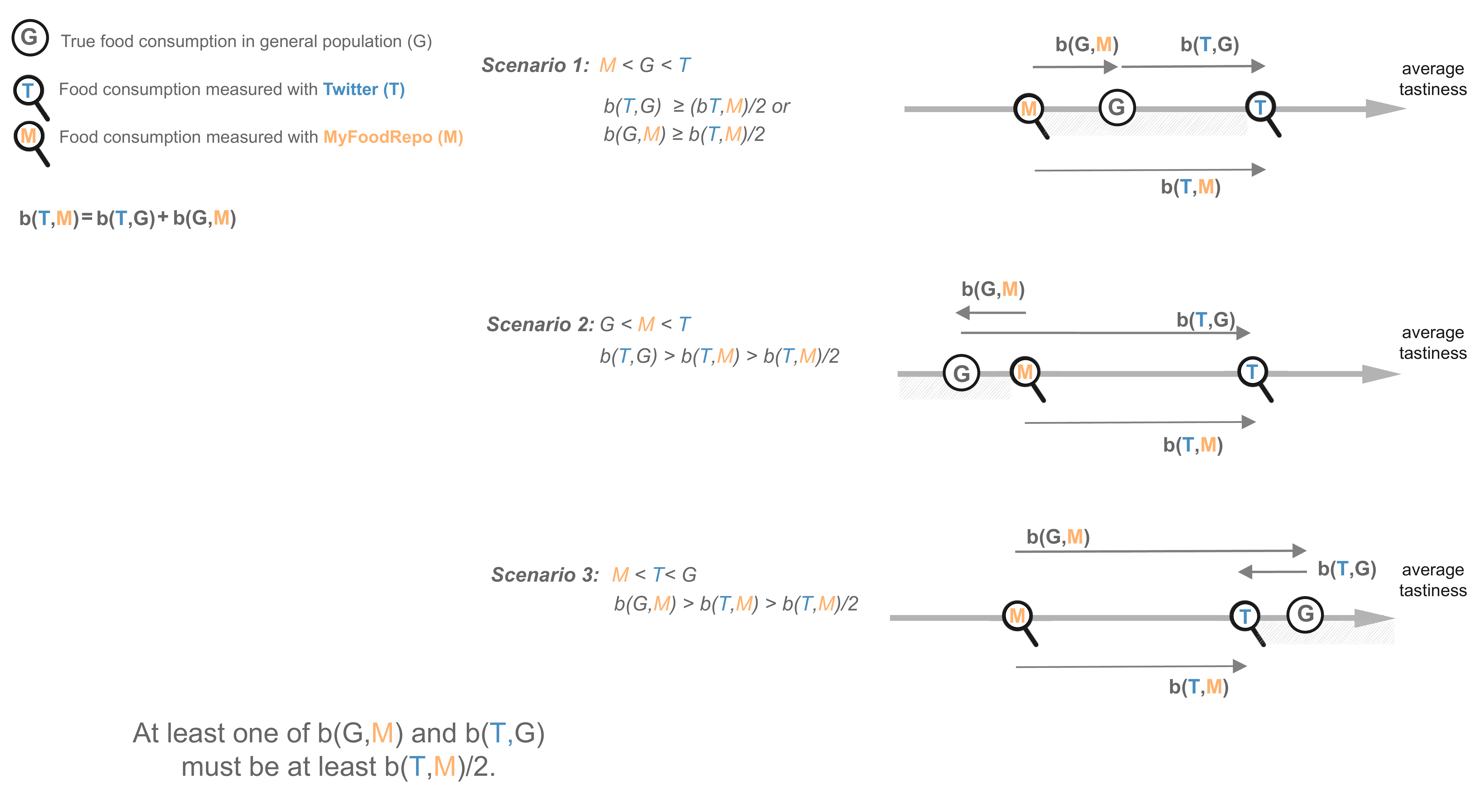}
    \caption{
    \boldmath
    \rev{\textbf{Illustration of (tastiness) biases between true food consumption $G$ in general population, tracked food $M$, and tweeted food $T$.}
    \unboldmath
    The bias between tweeted food and tracked food, $b(T,M)$, is characterized in our study, whereas $b(G,M)$ and $b(T,G)$ are unobserved. Although $b(T,G)$ and $b(G,M)$ are unobserved, at least one of $b(T,G)$ and $b(G,M)$ must be at least $b(T,M)/2$. The three illustrated possible scenarios depict situations where (1) $M < G < T$, (2) $ G < M < T$, or (3) $M < T < G$.}}
    \label{fig:biases}
\end{figure*}

\rev{For concreteness, consider tastiness (but the following argument equally applies to all other dimensions studied here).
Let $T$ denote the average tastiness score estimated via Twitter, $M$ the average tastiness score estimated via MyFoodRepo, and $G$ the true (unobserved) average tastiness score of food actually consumed by the general Swiss population.
As before (\cf\ \Eqnref{eq:2} and \Secref{sec:RQ2}), let the bias $b(T,M)=T - M$,} \rev{and analogously, $b(T,G) = T - G$ and $b(G,M) = G - M$.} \rev{Although $G$, $b(T,G)$, and $b(G,M)$ are not observed, we have:}
\begin{equation}
\begin{aligned}
        \rev{ b(T,G) + b(G,M)}  & \rev{ = (T - G) + (G - M) } \\
         & \rev{ = T - M} \\
         & \rev{ = b(T,M)}\label{eq:triangle}, 
\end{aligned}
\end{equation}
\rev{as illustrated in \Figref{fig:biases} for the case $T > M$ (without loss of generality; if $T < M$, we may simply make the argument about $b(M,T)$ instead of $b(T,M)$).}

\rev{The established substantial and significant bias $b(T,M)$ along the four studied dimensions (\Secref{sec:results}) therefore implies a lower bound on the unobserved biases, since at least one of $b(T,G)$ and $b(G,M)$ must be at least $b(T,M)/2$. In other words, along all four studied dimensions, either Twitter or MyFoodRepo differs by at least $b(T,M)/2$ from the general population. Dividing the measured biases $b(T,M)$ by two, we find that the lower bound on the bias still corresponds to significant gaps in how tasty, caloric, healthy, and likely to have been consumed at home the food is. Therefore, at least one of Twitter and MyFoodRepo foods significantly differs from the foods consumed by the general population. For example, the measured bias $b(T,M)$ is $0.52$ $[0.46, 0.56]$ for how tasty the food is (\Secref{sec:RQ2}). The lower bound with the shifted corresponding 95\% confidence intervals, $b(T,M)/2=0.26$ $[0.20,0.30]$, still corresponds to significant gaps in how tasty the food is. }

\rev{Consequently, the fact that there is a divergence between food consumption as measured via food tracking and as measured via social media implies that at least one of the two is not a faithful representation of the true food consumption in the general population concerning how healthy, tasty, caloric, and likely to have been consumed at home the food is. \Figref{fig:biases} illustrates possible scenarios in more detail. On the one hand, it might be that the true food consumption in the general population is somewhere between the food consumption as measured with MyFoodRepo, and the food consumption as measured with Twitter (\ie, $M < G < T$, Scenario 1). On the other hand, true food consumption in the general population might be skewed, and even more extreme than either MyFoodRepo (\ie, $G < M < T$, Scenario 2) or Twitter (\ie, $M < T < G$, Scenario 3).}

\rev{In principle, MyFoodRepo might be the main source of bias (\ie, $G$ might be closer to $T$ than to $M$) since individuals might not track all consumed foods, and food trackers are known to be situated and contextualized \cite{luo2019co,chung2019identifying}. Similarly, consumed foods logged with the MyFoodRepo app are likely not representative of the full Swiss population. People who log food with the tracking app have access to a smartphone with an Internet connection and care about their diets. However, we argue that it is less likely that MyFoodRepo is the main source of bias since the majority of MyFoodRepo food images are collected from volunteers enrolled in a digital cohort called Food \& You\footnote{\url{https://www.digitalepidemiologylab.org/projects/food-and-you}} who are instructed and reminded to provide images of their complete daily food intake. By design, the food present in these images was actually consumed, and omissions were discouraged. Therefore, it appears less likely that MyFoodRepo could misrepresent the true food consumption in the general population to such an extent that it would fully explain the measured biases $b(T,M)$.}

\rev{Nonetheless, given the absence of data about the general population and the fact that all the considered scenarios (\cf \Figref{fig:biases}) are not strictly impossible, we remain agnostic about the true source of bias. We argue that researchers should be attentive and aim to establish evidence of validity before using either social media or tracking apps as a proxy for true food consumption in the general population, since at least one of them differs by at least $b(T,M)/2$ from the general population. Future work should apply our framework for bias estimation to a representative sample of the overall population, with all food consumption recorded. At the present time, doing so remains challenging as the images logged with the tracking app by the volunteers are as good a peek onto actual plates as we can currently get.}

\subsection{Implications}

\subsubsection{\rev{Implications for research studying food tracking as a proxy for offline behaviors}} \minor{Based on our considerations of the two platforms (\ie, Twitter and MyFoodRepo), we argue that it is less likely that food tracking is the main source of bias when estimating food consumption in the general population. Nonetheless, researchers relying on food tracking should be attentive before implicitly assuming that the tracked consumption perfectly reflects the true consumption. Whenever possible, further contextualization of the tracked consumption data and investigation of alternative digital traces of the studied persons can be beneficial for examining the validity and establishing robustness. For instance, if there is a concern that users consume food systematically different from the logged food, future deployments should consider designing logging reminders and nudges within the tracking applications, targeted towards and specifically encouraging logging the true behaviors. Future research can also encourage users to assess the accuracy of logging through the tracking applications, for instance, by self\hyp reporting the overall perceived truthfulness.}

\subsubsection{\rev{Implications for social media research studying online traces as a proxy for offline behaviors}} \rev{Studies using passively collected digital traces as a proxy for real behaviors need to be valid in order to support public health research and have implications for the design of policies and interventions that can impact health outcomes.} Based on our findings, we now highlight major \rev{potential} pitfalls that can threaten the validity of such applications and provide actionable implications for overcoming them.

\emph{Actionable implication 1: Addressing over- and underrepresentation of food types.} If researchers were to estimate what foods the \rev{general population} consumes based on the number of tweets containing these foods, \rev{the estimates could be biased}. We suggest triangulating social media behaviors with known government statistics whenever these are publicly available. For example, although bread is underrepresented on Twitter, while burgers are overrepresented, \rev{compared with consumed and tracked foods} (\Figref{fig:categories}), researchers could---even without access to logs of actual food consumption---identify the implausibly high prevalence of burgers on social media compared to bread by examining publicly available statistics \cite{meyre2017agriculture}. For comparison, in 2019, the average Swiss consumed 89~kg of products based on grains, compared to roughly half as many kilograms of meat (48~kg). When aggregate country-wide statistics are available for calibration, social media can still be used as a sensor for spatially and temporally fine-grained analyses (\eg, by neighborhood or during holidays). When no such statistics are available, researchers might be able to calibrate their methods on populations where statistics are available and adjust the final estimates. Domain knowledge about the populations being studied can also help in alleviating some of these disproportions. One could consider knowledge about foods that studied populations consume in a social context, foods frequently consumed by visitors and tourists, or only during special periods or occasions, and avoid studies being impacted by such idiosyncrasies.

\emph{Actionable implication 2: Foods with very biased nutritional properties.} \rev{If researchers were to estimate nutritional properties of foods that the general population consumes based on the tweets containing images of foods, the estimates could be biased}. Researchers should be careful about bias that stems from certain foods that appear particularly less healthy and more caloric compared to consumed \rev{and tracked} food, such as chocolate, cheese, chicken, cake, and egg (although these tags might not necessarily represent exactly the same food types in populations beyond Switzerland). Researchers should also be aware of differences in portion sizes that likely mediate the difference in calories. We suggest detecting and examining images with an implausible amount of calories. For example, a single image might not be taken into account if the estimated amount of calories is not within reasonable bounds around the recommended daily 2,000--2,500 calories for an adult. 

Similarly, when training machine learning models with datasets obtained from social media \rev{and aiming to generalize to the general population}, samples should be adjusted such that the amount of calories more closely mirrors the amount of calories and portion sizes of real food. Otherwise, models trained on social media data to estimate the amount of calories will not make valid estimations outside of the context of exaggerated social media foods.

\emph{Actionable implication 3: Addressing systematic discrepancies in appearance.} \rev{If researchers were to estimate appearance of foods that the general population consumes based on the tweets containing images of foods, the estimates could be biased}. Foods that people consume \rev{and track} tend to appear significantly less tasty, simpler and less elaborate, prepared in different ways, and smaller in portion size, compared to tweeted food. These are challenging biases to overcome, as there is a need to use human annotation or computer vision models. Note, however, that the foods that are most biased in terms of nutritional properties and appearance are also precisely those that are overrepresented on Twitter (\Figref{fig:categories}). Therefore, adjusting the bias in the distribution of foods is likely to alleviate the bias in nutritional properties and appearance.

\subsubsection{\rev{Implications for social media research studying online traces per se}} \rev{Research studying online communities and online content not as a proxy for real behaviors but as a \emph{phenomenon per se} need not necessarily worry about validity issues and potential pitfalls. Such typical applications include studies characterizing online communities and specific users, such as users self\hyp reporting eating disorders online \cite{hunger2016,pro_eating2016} or online eating disorder support communities \cite{de2015anorexia,chancellor2016thyghgapp}. The behaviors of interest in these cases are precisely the online behaviors (\ie, the information that users choose to post). Similarly, studies developing machine learning models leveraging social media data that are \emph{not} concerned with performance generalization beyond the platform and to the general populations are not necessarily impacted by these biases. Such applications might include social media food recognition \cite{amato2017social,yanai2014twitter,sahoo2019foodai,bossard2014food} 
or learning online food image embeddings \cite{salvador2017learning}.}

\subsubsection{Implications for food representation and users' well\hyp being} Beyond the above implications for social media research, our results have implications for understanding the complex relationship between technology use and the well\hyp being of social media users. In the case of food, Twitter users are exposed to unrealistic mirrors of reality \cite{bail2021breaking}, since foods that people actually consume \rev{and track} are smaller, less ``fancy'', and less visually appealing (\Figref{fig:medians}, \Tabref{tab:words}). Such distortions might contribute to the high prevalence of social comparison \cite{frison2017browsing}, where, \eg, as much as one\hyp fifth of Facebook users can recall recently seeing a post that made them feel worse about themselves \cite{burke2020social}.
A user exposed to the social media portrayal of food might therefore believe that other people consume food that is tastier than the food they consume themselves. However, this would likely not be the case, due to the discrepancy between social media foods and consumed \rev{and tracked} foods. Social media might, in that case, promote an unhealthy relationship with food. Our findings have implications for research about the mechanisms of such social comparison.

\subsection{Limitations}

Next, we outline key limitations to be kept in mind when interpreting our results. In our main analyses, we study how foods are perceived by non\hyp expert crowd workers. The extent to which expert nutritionists would agree with such non\hyp experts is unknown. \rev{Furthermore, while the case study is focused on Switzerland, the crowdsourced workers are located in English-speaking countries, which might influence the food perception due to cultural factors.} Although the tags provided by the participants (\Tabref{tab:words}) provide insights about factors that guide their ratings, future work should more deeply investigate the nutritional properties of the studied foods in collaboration with expert annotators.

We note that the number of studied food images posted on Twitter---despite being the results of a best effort for completeness---is relatively small (around 3,700 Twitter photos of food, 2,400 of which were annotated). This number is small mostly due to the fact that we consider geolocated tweets only, and Switzerland is a relatively small country compared to the U.S., which has been studied in most related work \cite{mejova2015twitter,de2016characterizing,West:2013:CCI:2488388.2488510}.

We performed a case study of Twitter in Switzerland. \rev{Our findings cannot be assumed to generalize globally, and future work should apply our framework to other populations, other social media platforms and Web traces, and other food tracking apps.}
That said, this study may serve the purpose of a ``proof by counterexample'': \rev{we have identified one common setting where there is a divergence between food consumption as measured via food tracking and as measured via social media, implying that at least one of the two is not a faithful representation of the true food consumption in the general population. Hence, we should assume that there can be bias elsewhere, too. Researchers studying other populations should thus be attentive and aim to establish evidence of validity before using either social media or tracking apps as a proxy for the true food consumption in the general population. }

As a final limitation, we note that we did not include Swiss German dialect forms of keywords, since there is no written standard for Swiss German. 

\subsection{Potential sources of bias of social media \vs\ true food consumption}
We provide grounding and first insights about the validity of estimating food consumption from digital traces by \rev{contrasting consumed and tracked food with tweeted food}, controlling for location, period, and food types. Revealing exact mechanisms that can lead to the biases of social media traces \rev{as a proxy for true food consumption in the general population} is out of the scope of this work. However, in what follows, we consider potential sources of bias. We postulate that biases are driven by both {measurement error} (related to the operationalization of consumption via the concept of posting on Twitter) and {population error} (stemming from biases in subpopulations sharing food on Twitter) \cite{sen2021total}. Sources of \textit{measurement errors} might include these:

\begin{enumerate}
\item 
\emph{Construct validity.}  On the one hand, many foods that the Swiss consume are not posted on Twitter. Appealing food consumed in certain contexts is more likely to be shared, as positive and anticipated events are more likely to be disclosed on social media in general \cite{saha2021life}. Furthermore, photos published on Twitter may be self\hyp selected for higher quality, thus influencing how food is perceived by the annotators. On the other hand, not all foods shared on Twitter are necessarily consumed by the posting individual (especially not in their entirety, given the portion-size bias, \Tabref{tab:words}). Conceivably, certain tweets may originate from promotions, restaurants, or recipe sharing, all of which do not necessarily mirror actual consumption. In general, food images do not necessarily need to be related to consumption at all. They can mean something else entirely (\eg, a food can be a meme or a symbol of a political movement), although we did not find evidence of such biases in the studied data.
\item 
\emph{Platform effects.} Numerous applications and platforms support improving image quality and editing with filters, which can all contribute to the food image being more visually appealing and appearing tastier \cite{malik2016identifying}. 
\item 
\emph{Community feedback.} Feedback received from other platform members influences the type of dietary content which a social media user posts \cite{adelani2020estimating}, whereas negative feedback can lead to behavioral changes \cite{cheng2014community}. Biases in how food is represented online are implied by the design of online platforms. 
\end{enumerate}

Biases are also likely in part driven by \textit{population error.} Users of social media platforms do not mirror the general population, neither demographically nor regarding other attributes such as behaviors and interest \cite{lazer2021meaningful}. Users of public geotagged tweets are not randomly distributed over the general population \cite{malik2015population,fiesler2017or}. In the future, performing individual\hyp level studies, as opposed to the population\hyp level study reported here, will make it possible to disentangle measurement error from population error.


\subsection{Future work}

Beyond the already outlined future directions, the collected data can be used to further study patterns of sharing food online. Future work should further understand who shares food on Twitter (consumers, skilled individuals, but also non\hyp individual agents, such as restaurants or caterers). We expect that further characterizing user types would not change our main conclusions, but would reveal how biases vary between different strata of Twitter users. Future work should further study where they share it from (residential \vs commercial areas), when, and in what context, as well as what are the predictors of engagement with food on Twitter.

\subsection{Implications beyond food}

Researching human behaviors beyond food, our crowdsourcing framework can be used to measure many types of biases, including, but not limited to, 
politics and activism or behaviors important for health and well\hyp being, such as
fitness and time spent in nature,
travel, fashion and aesthetics, socialization, or
pet ownership.
Resolving the questions of truthfulness and validity of \rev{digital traces} beyond food is an important direction for future research. Moving forward, it is increasingly necessary for the our research community to invest in the collection of datasets closely mirroring real behaviors. 

\subsection{Code and data}
\label{sec:codeanddata}

Code and data necessary to reproduce our results at are publicly available at \url{https://github.com/epfl-dlab/biased-bytes}.

\section{Conclusion}
\label{sec:conclusion}

\rev{
As we conduct more research analyzing digital trace data, methods that offer insights into the validity of the new measures become increasingly necessary. Controlling for location, period and food types, we find that, overall, foods shared on social media significantly diverge from consumed and tracked foods. The fact that there is a divergence between food consumption as measured via food tracking and as measured via social media implies that at least one of the two is not a faithful representation of the true food consumption in the general population. Our study design lets us identify a lower bound: at least one of Twitter and MyFoodRepo diverges from the general Swiss population by at least half of the measured bias between the two platforms. We envision that the findings reported here will inform researchers in their efforts to study dietary behaviors. We also hope that the crowdsourcing framework for bias estimation and the initial quantifications will be useful broadly for research that leverages digital traces in the context of diets and beyond.
}

{\small
\section*{Acknowledgments}

We thank Maxime Peyrard for helpful feedback, and the Digital Epidemiology Lab for making the MyFoodRepo dataset publicly available via the AIcrowd Food Recognition Challenge. We acknowledge support from Microsoft (Swiss Joint Research Center),
Swiss National Science Foundation (grant 200021\_185043),
Collaborative Research on Science and Society (CROSS),
European Union (TAILOR, grant 952215),
Facebook,
and Google.
Finally, a hat tip to Marcel Salath\'e for having baked the top-rated pizza in the MyFoodRepo dataset (see \Figref{fig:diagram}).
}

\bibliographystyle{ACM-Reference-Format}
\bibliography{kristina_phd_thesis_bibliography}


\begin{thebibliography}{106}


\ifx \showCODEN    \undefined \def \showCODEN     #1{\unskip}     \fi
\ifx \showDOI      \undefined \def \showDOI       #1{#1}\fi
\ifx \showISBNx    \undefined \def \showISBNx     #1{\unskip}     \fi
\ifx \showISBNxiii \undefined \def \showISBNxiii  #1{\unskip}     \fi
\ifx \showISSN     \undefined \def \showISSN      #1{\unskip}     \fi
\ifx \showLCCN     \undefined \def \showLCCN      #1{\unskip}     \fi
\ifx \shownote     \undefined \def \shownote      #1{#1}          \fi
\ifx \showarticletitle \undefined \def \showarticletitle #1{#1}   \fi
\ifx \showURL      \undefined \def \showURL       {\relax}        \fi
\providecommand\bibfield[2]{#2}
\providecommand\bibinfo[2]{#2}
\providecommand\natexlab[1]{#1}
\providecommand\showeprint[2][]{arXiv:#2}

\bibitem[Abbar et~al\mbox{.}(2018)]%
        {10.1145/3209542.3209575}
\bibfield{author}{\bibinfo{person}{Sofiane Abbar}, \bibinfo{person}{Carlos
  Castillo}, {and} \bibinfo{person}{Antonio Sanfilippo}.}
  \bibinfo{year}{2018}\natexlab{}.
\newblock \showarticletitle{To Post or Not to Post: Using Online Trends to
  Predict Popularity of Offline Content}. In
  \bibinfo{booktitle}{\emph{Proceedings of the 29th Conference on Hypertext and
  Social Media}}.
\newblock


\bibitem[Abbar et~al\mbox{.}(2015)]%
        {abbar2015you}
\bibfield{author}{\bibinfo{person}{Sofiane Abbar}, \bibinfo{person}{Yelena
  Mejova}, {and} \bibinfo{person}{Ingmar Weber}.}
  \bibinfo{year}{2015}\natexlab{}.
\newblock \showarticletitle{You tweet what you eat: Studying food consumption
  through Twitter}. In \bibinfo{booktitle}{\emph{Proc. of the 33rd Conference
  on Human Factors in Computing Systems (CHI)}}.
\newblock


\bibitem[Achananuparp and Weber(2016)]%
        {achananuparp2016extracting}
\bibfield{author}{\bibinfo{person}{Palakorn Achananuparp} {and}
  \bibinfo{person}{Ingmar Weber}.} \bibinfo{year}{2016}\natexlab{}.
\newblock \showarticletitle{Extracting food substitutes from food diary via
  distributional similarity}.
\newblock  (\bibinfo{year}{2016}).
\newblock


\bibitem[Adelani et~al\mbox{.}(2020)]%
        {adelani2020estimating}
\bibfield{author}{\bibinfo{person}{David~Ifeoluwa Adelani},
  \bibinfo{person}{Ryota Kobayashi}, \bibinfo{person}{Ingmar Weber}, {and}
  \bibinfo{person}{Przemyslaw~A Grabowicz}.} \bibinfo{year}{2020}\natexlab{}.
\newblock \showarticletitle{Estimating community feedback effect on topic
  choice in social media with predictive modeling}.
\newblock \bibinfo{journal}{\emph{EPJ Data Science}} \bibinfo{volume}{9},
  \bibinfo{number}{1} (\bibinfo{year}{2020}).
\newblock


\bibitem[Agresti(2003)]%
        {agresti2003categorical}
\bibfield{author}{\bibinfo{person}{Alan Agresti}.}
  \bibinfo{year}{2003}\natexlab{}.
\newblock \bibinfo{booktitle}{\emph{Categorical data analysis}}.
  Vol.~\bibinfo{volume}{482}.
\newblock


\bibitem[Aiello et~al\mbox{.}(2020)]%
        {aiello2020tesco}
\bibfield{author}{\bibinfo{person}{Luca~Maria Aiello}, \bibinfo{person}{Daniele
  Quercia}, \bibinfo{person}{Rossano Schifanella}, {and} \bibinfo{person}{Lucia
  Del~Prete}.} \bibinfo{year}{2020}\natexlab{}.
\newblock \showarticletitle{Tesco Grocery 1.0, a large-scale dataset of grocery
  purchases in London}.
\newblock \bibinfo{journal}{\emph{Scientific Data}} \bibinfo{volume}{7},
  \bibinfo{number}{1} (\bibinfo{year}{2020}).
\newblock


\bibitem[Aiello et~al\mbox{.}(2019)]%
        {aiello2019large}
\bibfield{author}{\bibinfo{person}{Luca~Maria Aiello}, \bibinfo{person}{Rossano
  Schifanella}, \bibinfo{person}{Daniele Quercia}, {and} \bibinfo{person}{Lucia
  Del~Prete}.} \bibinfo{year}{2019}\natexlab{}.
\newblock \showarticletitle{Large-scale and high-resolution analysis of food
  purchases and health outcomes}.
\newblock \bibinfo{journal}{\emph{EPJ Data Science}} \bibinfo{volume}{8},
  \bibinfo{number}{1} (\bibinfo{year}{2019}).
\newblock


\bibitem[Althoff et~al\mbox{.}(2017)]%
        {althoff2017online}
\bibfield{author}{\bibinfo{person}{Tim Althoff}, \bibinfo{person}{Pranav
  Jindal}, {and} \bibinfo{person}{Jure Leskovec}.}
  \bibinfo{year}{2017}\natexlab{}.
\newblock \showarticletitle{Online actions with offline impact: How online
  social networks influence online and offline user behavior}. In
  \bibinfo{booktitle}{\emph{Proc. of the 10th ACM International Conference on
  Web Search and Data Mining (WSDM)}}.
\newblock


\bibitem[Amato et~al\mbox{.}(2017)]%
        {amato2017social}
\bibfield{author}{\bibinfo{person}{Giuseppe Amato}, \bibinfo{person}{Paolo
  Bolettieri}, \bibinfo{person}{Vinicius Monteiro~de Lira},
  \bibinfo{person}{Cristina~Ioana Muntean}, \bibinfo{person}{Raffaele Perego},
  {and} \bibinfo{person}{Chiara Renso}.} \bibinfo{year}{2017}\natexlab{}.
\newblock \showarticletitle{Social media image recognition for food trend
  analysis}. In \bibinfo{booktitle}{\emph{Proc. of the 40th International ACM
  Conference on Research and Development in Information Retrieval (SIGIR)}}.
\newblock


\bibitem[Bail(2021)]%
        {bail2021breaking}
\bibfield{author}{\bibinfo{person}{Chris Bail}.}
  \bibinfo{year}{2021}\natexlab{}.
\newblock \bibinfo{booktitle}{\emph{Breaking the Social Media Prism}}.
\newblock


\bibitem[Biel et~al\mbox{.}(2018)]%
        {biel2018bites}
\bibfield{author}{\bibinfo{person}{Joan-Isaac Biel}, \bibinfo{person}{Nathalie
  Martin}, \bibinfo{person}{David Labbe}, {and} \bibinfo{person}{Daniel
  Gatica-Perez}.} \bibinfo{year}{2018}\natexlab{}.
\newblock \showarticletitle{Bites ‘n’bits: Inferring eating behavior from
  contextual mobile data}.
\newblock \bibinfo{journal}{\emph{Proceedings of the ACM on Interactive,
  Mobile, Wearable and Ubiquitous Technologies}} \bibinfo{volume}{1},
  \bibinfo{number}{4} (\bibinfo{year}{2018}).
\newblock


\bibitem[Bin~Morshed et~al\mbox{.}(2020)]%
        {info:doi/10.2196/20625}
\bibfield{author}{\bibinfo{person}{Mehrab Bin~Morshed},
  \bibinfo{person}{Samruddhi~Shreeram Kulkarni}, \bibinfo{person}{Richard Li},
  \bibinfo{person}{Koustuv Saha}, \bibinfo{person}{Leah~Galante Roper},
  \bibinfo{person}{Lama Nachman}, \bibinfo{person}{Hong Lu},
  \bibinfo{person}{Lucia Mirabella}, \bibinfo{person}{Sanjeev Srivastava},
  \bibinfo{person}{Munmun De~Choudhury}, \bibinfo{person}{Kaya de Barbaro},
  \bibinfo{person}{Thomas Ploetz}, {and} \bibinfo{person}{Gregory~D Abowd}.}
  \bibinfo{year}{2020}\natexlab{}.
\newblock \showarticletitle{A Real-Time eating detection system for capturing
  eating moments and triggering ecological momentary assessments to obtain
  further context: system development and validation study}.
\newblock \bibinfo{journal}{\emph{JMIR mHealth and uHealth}}
  \bibinfo{volume}{8}, \bibinfo{number}{12} (\bibinfo{year}{2020}).
\newblock


\bibitem[Blumenstock et~al\mbox{.}(2015)]%
        {blumenstock2015predicting}
\bibfield{author}{\bibinfo{person}{Joshua Blumenstock},
  \bibinfo{person}{Gabriel Cadamuro}, {and} \bibinfo{person}{Robert On}.}
  \bibinfo{year}{2015}\natexlab{}.
\newblock \showarticletitle{Predicting poverty and wealth from mobile phone
  metadata}.
\newblock \bibinfo{journal}{\emph{Science}} \bibinfo{volume}{350},
  \bibinfo{number}{6264} (\bibinfo{year}{2015}).
\newblock


\bibitem[Bossard et~al\mbox{.}(2014)]%
        {bossard2014food}
\bibfield{author}{\bibinfo{person}{Lukas Bossard}, \bibinfo{person}{Matthieu
  Guillaumin}, {and} \bibinfo{person}{Luc Van~Gool}.}
  \bibinfo{year}{2014}\natexlab{}.
\newblock \showarticletitle{Food-101--mining discriminative components with
  random forests}. In \bibinfo{booktitle}{\emph{European Conference on Computer
  Vision}}. Springer.
\newblock


\bibitem[Bowling(2005)]%
        {bowling2005mode}
\bibfield{author}{\bibinfo{person}{Ann Bowling}.}
  \bibinfo{year}{2005}\natexlab{}.
\newblock \showarticletitle{Mode of questionnaire administration can have
  serious effects on data quality}.
\newblock \bibinfo{journal}{\emph{Journal of Public Health}}
  \bibinfo{volume}{27}, \bibinfo{number}{3} (\bibinfo{year}{2005}).
\newblock


\bibitem[Bradley and Terry(1952)]%
        {bradley1952rank}
\bibfield{author}{\bibinfo{person}{Ralph~Allan Bradley} {and}
  \bibinfo{person}{Milton~E Terry}.} \bibinfo{year}{1952}\natexlab{}.
\newblock \showarticletitle{Rank analysis of incomplete block designs: I. The
  method of paired comparisons}.
\newblock \bibinfo{journal}{\emph{Biometrika}} \bibinfo{volume}{39},
  \bibinfo{number}{3/4} (\bibinfo{year}{1952}).
\newblock


\bibitem[Buckeridge et~al\mbox{.}(2014)]%
        {buckeridge2014method}
\bibfield{author}{\bibinfo{person}{David~L Buckeridge}, \bibinfo{person}{Katia
  Charland}, \bibinfo{person}{Alice Labban}, {and} \bibinfo{person}{Yu Ma}.}
  \bibinfo{year}{2014}\natexlab{}.
\newblock \showarticletitle{A method for neighborhood level surveillance of
  food purchasing}.
\newblock \bibinfo{journal}{\emph{Annals of the New York Academy of Sciences}}
  \bibinfo{volume}{1331}, \bibinfo{number}{1} (\bibinfo{year}{2014}).
\newblock


\bibitem[Burke et~al\mbox{.}(2020)]%
        {burke2020social}
\bibfield{author}{\bibinfo{person}{Moira Burke}, \bibinfo{person}{Justin
  Cheng}, {and} \bibinfo{person}{Bethany de Gant}.}
  \bibinfo{year}{2020}\natexlab{}.
\newblock \showarticletitle{Social comparison and Facebook: Feedback,
  positivity, and opportunities for comparison}. In
  \bibinfo{booktitle}{\emph{Proceedings of the 2020 Conference on Human Factors
  in Computing Systems (CHI)}}.
\newblock


\bibitem[Chancellor et~al\mbox{.}(2016a)]%
        {pro_eating2016}
\bibfield{author}{\bibinfo{person}{Stevie Chancellor}, \bibinfo{person}{Zhiyuan
  Lin}, \bibinfo{person}{Erica~L. Goodman}, \bibinfo{person}{Stephanie Zerwas},
  {and} \bibinfo{person}{Munmun De~Choudhury}.}
  \bibinfo{year}{2016}\natexlab{a}.
\newblock \showarticletitle{Quantifying and predicting mental illness severity
  in online pro-eating disorder communities}. In
  \bibinfo{booktitle}{\emph{Proc. of the 2016 ACM Conference on Computer
  Supported Cooperative Work and Social Computing (CSCW)}}.
\newblock


\bibitem[Chancellor et~al\mbox{.}(2016b)]%
        {chancellor2016thyghgapp}
\bibfield{author}{\bibinfo{person}{Stevie Chancellor},
  \bibinfo{person}{Jessica~Annette Pater}, \bibinfo{person}{Trustin Clear},
  \bibinfo{person}{Eric Gilbert}, {and} \bibinfo{person}{Munmun De~Choudhury}.}
  \bibinfo{year}{2016}\natexlab{b}.
\newblock \showarticletitle{\# thyghgapp: Instagram content moderation and
  lexical variation in pro-eating disorder communities}. In
  \bibinfo{booktitle}{\emph{Proc. of the 2016 ACM Conference on Computer
  Supported Cooperative Work and Social Computing (CSCW)}}.
\newblock


\bibitem[Chen et~al\mbox{.}(2013)]%
        {chen2013pairwise}
\bibfield{author}{\bibinfo{person}{Xi Chen}, \bibinfo{person}{Paul~N Bennett},
  \bibinfo{person}{Kevyn Collins-Thompson}, {and} \bibinfo{person}{Eric
  Horvitz}.} \bibinfo{year}{2013}\natexlab{}.
\newblock \showarticletitle{Pairwise ranking aggregation in a crowdsourced
  setting}. In \bibinfo{booktitle}{\emph{Proc. of the 6th ACM International
  Conference on Web Search and Data Mining (WSDM)}}.
\newblock


\bibitem[Cheng et~al\mbox{.}(2014)]%
        {cheng2014community}
\bibfield{author}{\bibinfo{person}{Justin Cheng}, \bibinfo{person}{Cristian
  Danescu-Niculescu-Mizil}, {and} \bibinfo{person}{Jure Leskovec}.}
  \bibinfo{year}{2014}\natexlab{}.
\newblock \showarticletitle{How community feedback shapes user behavior}. In
  \bibinfo{booktitle}{\emph{Proc. of the eighth International AAAI Conference
  on Weblogs and Social Media (ICWSM)}}.
\newblock


\bibitem[Chorley et~al\mbox{.}(2016)]%
        {chorley2016pub}
\bibfield{author}{\bibinfo{person}{Martin~J Chorley}, \bibinfo{person}{Luca
  Rossi}, \bibinfo{person}{Gareth Tyson}, {and} \bibinfo{person}{Matthew~J
  Williams}.} \bibinfo{year}{2016}\natexlab{}.
\newblock \showarticletitle{Pub crawling at scale: tapping untappd to explore
  social drinking}. In \bibinfo{booktitle}{\emph{Proc. of the 10th
  International AAAI Conference on Web and Social Media (ICWSM)}}.
\newblock


\bibitem[Christakis and Fowler(2007)]%
        {christakis2007spread}
\bibfield{author}{\bibinfo{person}{Nicholas~A Christakis} {and}
  \bibinfo{person}{James~H Fowler}.} \bibinfo{year}{2007}\natexlab{}.
\newblock \showarticletitle{The spread of obesity in a large social network
  over 32 years}.
\newblock \bibinfo{journal}{\emph{New England Journal of Medicine (NEJM)}}
  \bibinfo{volume}{357}, \bibinfo{number}{4} (\bibinfo{year}{2007}).
\newblock


\bibitem[Chung et~al\mbox{.}(2017)]%
        {chung2017personal}
\bibfield{author}{\bibinfo{person}{Chia-Fang Chung}, \bibinfo{person}{Elena
  Agapie}, \bibinfo{person}{Jessica Schroeder}, \bibinfo{person}{Sonali
  Mishra}, \bibinfo{person}{James Fogarty}, {and} \bibinfo{person}{Sean~A
  Munson}.} \bibinfo{year}{2017}\natexlab{}.
\newblock \showarticletitle{When personal tracking becomes social: Examining
  the use of Instagram for healthy eating}. In \bibinfo{booktitle}{\emph{Proc.
  of the 2017 Conference on Human Factors in Computing Systems (CHI)}}.
\newblock


\bibitem[Chung et~al\mbox{.}(2015)]%
        {chung2015more}
\bibfield{author}{\bibinfo{person}{Chia-Fang Chung}, \bibinfo{person}{Jonathan
  Cook}, \bibinfo{person}{Elizabeth Bales}, \bibinfo{person}{Jasmine Zia},
  {and} \bibinfo{person}{Sean~A Munson}.} \bibinfo{year}{2015}\natexlab{}.
\newblock \showarticletitle{More than telemonitoring: health provider use and
  nonuse of life-log data in irritable bowel syndrome and weight management}.
\newblock \bibinfo{journal}{\emph{Journal of Medical Internet Research}}
  \bibinfo{volume}{17}, \bibinfo{number}{8} (\bibinfo{year}{2015}).
\newblock


\bibitem[Chung et~al\mbox{.}(2019)]%
        {chung2019identifying}
\bibfield{author}{\bibinfo{person}{Chia-Fang Chung}, \bibinfo{person}{Qiaosi
  Wang}, \bibinfo{person}{Jessica Schroeder}, \bibinfo{person}{Allison Cole},
  \bibinfo{person}{Jasmine Zia}, \bibinfo{person}{James Fogarty}, {and}
  \bibinfo{person}{Sean~A Munson}.} \bibinfo{year}{2019}\natexlab{}.
\newblock \showarticletitle{Identifying and planning for individualized change:
  Patient-provider collaboration using lightweight food diaries in healthy
  eating and irritable bowel syndrome}.
\newblock \bibinfo{journal}{\emph{Proceedings of the ACM on Interactive,
  Mobile, Wearable and Ubiquitous Technologies (IMWUT)}} \bibinfo{volume}{3},
  \bibinfo{number}{1} (\bibinfo{year}{2019}).
\newblock


\bibitem[Corbett-Davies et~al\mbox{.}(2019)]%
        {corbett2017algorithmic}
\bibfield{author}{\bibinfo{person}{Sam Corbett-Davies}, \bibinfo{person}{Emma
  Pierson}, \bibinfo{person}{Avi Feller}, \bibinfo{person}{Sharad Goel}, {and}
  \bibinfo{person}{Aziz Huq}.} \bibinfo{year}{2019}\natexlab{}.
\newblock \showarticletitle{Algorithmic decision making and the cost of
  fairness}. In \bibinfo{booktitle}{\emph{Proceedings of the 23rd ACM
  International Conference on Knowledge Discovery and Data Mining (KDD)}}.
\newblock


\bibitem[Cordeiro et~al\mbox{.}(2015a)]%
        {food_journal2015}
\bibfield{author}{\bibinfo{person}{Felicia Cordeiro},
  \bibinfo{person}{Elizabeth Bales}, \bibinfo{person}{Erin Cherry}, {and}
  \bibinfo{person}{James Fogarty}.} \bibinfo{year}{2015}\natexlab{a}.
\newblock \showarticletitle{Rethinking the mobile food journal: Exploring
  opportunities for lightweight photo-based capture}. In
  \bibinfo{booktitle}{\emph{Proc. of the 2015 Conference on Human Factors in
  Computing Systems (CHI)}}.
\newblock


\bibitem[Cordeiro et~al\mbox{.}(2015b)]%
        {barriers_negative2015}
\bibfield{author}{\bibinfo{person}{Felicia Cordeiro},
  \bibinfo{person}{Daniel~A. Epstein}, \bibinfo{person}{Edison Thomaz},
  \bibinfo{person}{Elizabeth Bales}, \bibinfo{person}{Arvind~K. Jagannathan},
  \bibinfo{person}{Gregory~D. Abowd}, {and} \bibinfo{person}{James Fogarty}.}
  \bibinfo{year}{2015}\natexlab{b}.
\newblock \showarticletitle{Barriers and negative nudges: Exploring challenges
  in food journaling}. In \bibinfo{booktitle}{\emph{Proc. of the 2015
  Conference on Human Factors in Computing Systems (CHI)}}.
\newblock


\bibitem[Danescu-Niculescu-Mizil et~al\mbox{.}(2013)]%
        {danescu2013no}
\bibfield{author}{\bibinfo{person}{Cristian Danescu-Niculescu-Mizil},
  \bibinfo{person}{Robert West}, \bibinfo{person}{Dan Jurafsky},
  \bibinfo{person}{Jure Leskovec}, {and} \bibinfo{person}{Christopher Potts}.}
  \bibinfo{year}{2013}\natexlab{}.
\newblock \showarticletitle{No country for old members: User lifecycle and
  linguistic change in online communities}. In \bibinfo{booktitle}{\emph{Proc.
  of the 22nd International Conference on World Wide Web (TheWebConf)}}.
\newblock


\bibitem[De~Choudhury(2015)]%
        {de2015anorexia}
\bibfield{author}{\bibinfo{person}{Munmun De~Choudhury}.}
  \bibinfo{year}{2015}\natexlab{}.
\newblock \showarticletitle{Anorexia on tumblr: A characterization study}. In
  \bibinfo{booktitle}{\emph{Proc. of the 5th International Conference on
  Digital Health}}.
\newblock


\bibitem[De~Choudhury et~al\mbox{.}(2014)]%
        {de2014seeking}
\bibfield{author}{\bibinfo{person}{Munmun De~Choudhury},
  \bibinfo{person}{Meredith~Ringel Morris}, {and} \bibinfo{person}{Ryen~W
  White}.} \bibinfo{year}{2014}\natexlab{}.
\newblock \showarticletitle{Seeking and sharing health information online:
  Comparing search engines and social media}. In
  \bibinfo{booktitle}{\emph{Proc. of the 2014 Conference on Human Factors in
  Computing Systems (CHI)}}.
\newblock


\bibitem[De~Choudhury et~al\mbox{.}(2016)]%
        {de2016characterizing}
\bibfield{author}{\bibinfo{person}{Munmun De~Choudhury},
  \bibinfo{person}{Sanket Sharma}, {and} \bibinfo{person}{Emre K{\i}c{\i}man}.}
  \bibinfo{year}{2016}\natexlab{}.
\newblock \showarticletitle{Characterizing dietary choices, nutrition, and
  language in food deserts via social media}. In
  \bibinfo{booktitle}{\emph{Proc. of the 2016 ACM Conference on Computer
  Supported Cooperative Work and Social Computing (CSCW)}}.
\newblock


\bibitem[Dijkstra(2022)]%
        {dijkstra2022cross}
\bibfield{author}{\bibinfo{person}{Neele Dijkstra}.}
  \bibinfo{year}{2022}\natexlab{}.
\newblock \emph{\bibinfo{title}{Cross-modal recipe analysis for fine-grained
  geographical mapping of food consumption from images and supermarket sales}}.
\newblock \bibinfo{thesistype}{Master's\ thesis}.
\newblock


\bibitem[Dunford et~al\mbox{.}(2014)]%
        {dunford2014foodswitch}
\bibfield{author}{\bibinfo{person}{Elizabeth Dunford}, \bibinfo{person}{Helen
  Trevena}, \bibinfo{person}{Chester Goodsell}, \bibinfo{person}{Ka~Hung Ng},
  \bibinfo{person}{Jacqui Webster}, \bibinfo{person}{Audra Millis},
  \bibinfo{person}{Stan Goldstein}, \bibinfo{person}{Orla Hugueniot}, {and}
  \bibinfo{person}{Bruce Neal}.} \bibinfo{year}{2014}\natexlab{}.
\newblock \showarticletitle{FoodSwitch: a mobile phone app to enable consumers
  to make healthier food choices and crowdsourcing of national food composition
  data}.
\newblock \bibinfo{journal}{\emph{JMIR mHealth and uHealth}}
  (\bibinfo{year}{2014}).
\newblock


\bibitem[El~Fatouhi et~al\mbox{.}(2022)]%
        {el2022associations}
\bibfield{author}{\bibinfo{person}{Douae El~Fatouhi}, \bibinfo{person}{Harris
  H{\'e}ritier}, \bibinfo{person}{Chlo{\'e} All{\'e}mann},
  \bibinfo{person}{Laurent Malisoux}, \bibinfo{person}{Nasser Laouali},
  \bibinfo{person}{Jean-Pierre Riveline}, \bibinfo{person}{Marcel Salath{\'e}},
  {and} \bibinfo{person}{Guy Fagherazzi}.} \bibinfo{year}{2022}\natexlab{}.
\newblock \showarticletitle{Associations Between Device-Measured Physical
  Activity and Glycemic Control and Variability Indices Under Free-Living
  Conditions}.
\newblock \bibinfo{journal}{\emph{Diabetes Technology \& Therapeutics}}
  \bibinfo{volume}{24}, \bibinfo{number}{3} (\bibinfo{year}{2022}).
\newblock


\bibitem[Fiesler et~al\mbox{.}(2017)]%
        {fiesler2017or}
\bibfield{author}{\bibinfo{person}{Casey Fiesler}, \bibinfo{person}{Michaelanne
  Dye}, \bibinfo{person}{Jessica~L Feuston}, \bibinfo{person}{Chaya
  Hiruncharoenvate}, \bibinfo{person}{Clayton~J Hutto},
  \bibinfo{person}{Shannon Morrison}, \bibinfo{person}{Parisa
  Khanipour~Roshan}, \bibinfo{person}{Umashanthi Pavalanathan},
  \bibinfo{person}{Amy~S Bruckman}, \bibinfo{person}{Munmun De~Choudhury},
  {et~al\mbox{.}}} \bibinfo{year}{2017}\natexlab{}.
\newblock \showarticletitle{What (or who) is public? Privacy settings and
  social media content sharing}. In \bibinfo{booktitle}{\emph{Proc. of the 2017
  ACM Conference on Computer Supported Cooperative Work and Social Computing
  (CSCW)}}.
\newblock


\bibitem[Freitas et~al\mbox{.}(2020)]%
        {freitas2020myfood}
\bibfield{author}{\bibinfo{person}{Charles~NC Freitas},
  \bibinfo{person}{Filipe~R Cordeiro}, {and} \bibinfo{person}{Valmir Macario}.}
  \bibinfo{year}{2020}\natexlab{}.
\newblock \showarticletitle{Myfood: A food segmentation and classification
  system to aid nutritional monitoring}. In \bibinfo{booktitle}{\emph{2020 33rd
  SIBGRAPI Conference on Graphics, Patterns and Images (SIBGRAPI)}}.
\newblock


\bibitem[Frison and Eggermont(2017)]%
        {frison2017browsing}
\bibfield{author}{\bibinfo{person}{Eline Frison} {and} \bibinfo{person}{Steven
  Eggermont}.} \bibinfo{year}{2017}\natexlab{}.
\newblock \showarticletitle{Browsing, posting, and liking on Instagram: The
  reciprocal relationships between different types of Instagram use and
  adolescents' depressed mood}.
\newblock \bibinfo{journal}{\emph{Cyberpsychology, Behavior, and Social
  Networking}} \bibinfo{volume}{20}, \bibinfo{number}{10}
  (\bibinfo{year}{2017}).
\newblock


\bibitem[Garimella et~al\mbox{.}(2016)]%
        {garimella2016social}
\bibfield{author}{\bibinfo{person}{Venkata Rama~Kiran Garimella},
  \bibinfo{person}{Abdulrahman Alfayad}, {and} \bibinfo{person}{Ingmar Weber}.}
  \bibinfo{year}{2016}\natexlab{}.
\newblock \showarticletitle{Social media image analysis for public health}. In
  \bibinfo{booktitle}{\emph{Proc. of the 2016 Conference on Human Factors in
  Computing Systems (CHI)}}.
\newblock


\bibitem[Gatica-Perez et~al\mbox{.}(2019)]%
        {gatica2019discovering}
\bibfield{author}{\bibinfo{person}{Daniel Gatica-Perez},
  \bibinfo{person}{Joan-Isaac Biel}, \bibinfo{person}{David Labbe}, {and}
  \bibinfo{person}{Nathalie Martin}.} \bibinfo{year}{2019}\natexlab{}.
\newblock \showarticletitle{Discovering eating routines in context with a
  smartphone app}. In \bibinfo{booktitle}{\emph{Adjunct Proceedings of the 2019
  ACM International Joint Conference on Pervasive and Ubiquitous Computing and
  Proceedings of the 2019 ACM International Symposium on Wearable Computers}}.
\newblock


\bibitem[Gligori{\'c} et~al\mbox{.}(2022)]%
        {gligoric2022population}
\bibfield{author}{\bibinfo{person}{Kristina Gligori{\'c}},
  \bibinfo{person}{Arnaud Chiolero}, \bibinfo{person}{Emre K{\i}c{\i}man},
  \bibinfo{person}{Ryen~W White}, {and} \bibinfo{person}{Robert West}.}
  \bibinfo{year}{2022}\natexlab{}.
\newblock \showarticletitle{Population-scale dietary interests during the
  COVID-19 pandemic}.
\newblock \bibinfo{journal}{\emph{Nature Communications}} \bibinfo{volume}{13},
  \bibinfo{number}{1} (\bibinfo{year}{2022}).
\newblock


\bibitem[Gligori{\'c} et~al\mbox{.}(2021)]%
        {gligoric2021ties}
\bibfield{author}{\bibinfo{person}{Kristina Gligori{\'c}},
  \bibinfo{person}{Ryen~W. White}, \bibinfo{person}{Emre Kiciman},
  \bibinfo{person}{Eric Horvitz}, \bibinfo{person}{Arnaud Chiolero}, {and}
  \bibinfo{person}{Robert West}.} \bibinfo{year}{2021}\natexlab{}.
\newblock \showarticletitle{Formation of Social Ties Influences Food Choice: A
  Campus-Wide Longitudinal Study}.
\newblock \bibinfo{journal}{\emph{Proc. of the ACM Conference on Computer
  Supported Cooperative Work and Social Computing (CSCW)}}
  (\bibinfo{year}{2021}).
\newblock


\bibitem[Gudivada et~al\mbox{.}(2015)]%
        {gudivada2015big}
\bibfield{author}{\bibinfo{person}{Venkat~N Gudivada}, \bibinfo{person}{Ricardo
  Baeza-Yates}, {and} \bibinfo{person}{Vijay~V Raghavan}.}
  \bibinfo{year}{2015}\natexlab{}.
\newblock \showarticletitle{Big data: Promises and problems}.
\newblock \bibinfo{journal}{\emph{Computer}} \bibinfo{volume}{48},
  \bibinfo{number}{03} (\bibinfo{year}{2015}).
\newblock


\bibitem[Habib and Saha(2010)]%
        {habib2010burden}
\bibfield{author}{\bibinfo{person}{Samira~Humaira Habib} {and}
  \bibinfo{person}{Soma Saha}.} \bibinfo{year}{2010}\natexlab{}.
\newblock \showarticletitle{Burden of non-communicable disease: global
  overview}.
\newblock \bibinfo{journal}{\emph{Diabetes \& Metabolic Syndrome: Clinical
  Research \& Reviews}} \bibinfo{volume}{4}, \bibinfo{number}{1}
  (\bibinfo{year}{2010}).
\newblock


\bibitem[Hann{\'a}k et~al\mbox{.}(2017)]%
        {hannak2017bias}
\bibfield{author}{\bibinfo{person}{Anik{\'o} Hann{\'a}k},
  \bibinfo{person}{Claudia Wagner}, \bibinfo{person}{David Garcia},
  \bibinfo{person}{Alan Mislove}, \bibinfo{person}{Markus Strohmaier}, {and}
  \bibinfo{person}{Christo Wilson}.} \bibinfo{year}{2017}\natexlab{}.
\newblock \showarticletitle{Bias in online freelance marketplaces: Evidence
  from taskrabbit and fiverr}. In \bibinfo{booktitle}{\emph{Proc. of the 2017
  ACM Conference on Computer Supported Cooperative Work and Social Computing
  (CSCW)}}.
\newblock


\bibitem[Harris et~al\mbox{.}(2012)]%
        {harris2012us}
\bibfield{author}{\bibinfo{person}{Jennifer~L Harris}, \bibinfo{person}{Sarah~E
  Speers}, \bibinfo{person}{Marlene~B Schwartz}, {and} \bibinfo{person}{Kelly~D
  Brownell}.} \bibinfo{year}{2012}\natexlab{}.
\newblock \showarticletitle{US food company branded advergames on the Internet:
  Children's exposure and effects on snack consumption}.
\newblock \bibinfo{journal}{\emph{Journal of Children and Media}}
  \bibinfo{volume}{6}, \bibinfo{number}{1} (\bibinfo{year}{2012}).
\newblock


\bibitem[He et~al\mbox{.}(2014)]%
        {he2016deep}
\bibfield{author}{\bibinfo{person}{Kaiming He}, \bibinfo{person}{Xiangyu
  Zhang}, \bibinfo{person}{Shaoqing Ren}, {and} \bibinfo{person}{Jian Sun}.}
  \bibinfo{year}{2014}\natexlab{}.
\newblock \showarticletitle{Deep residual learning for image recognition}. In
  \bibinfo{booktitle}{\emph{Proc. of the IEEE Conference on Computer Vision and
  Pattern Recognition (CVPR)}}.
\newblock


\bibitem[Hofman et~al\mbox{.}(2021)]%
        {hofman2021integrating}
\bibfield{author}{\bibinfo{person}{Jake~M Hofman}, \bibinfo{person}{Duncan~J
  Watts}, \bibinfo{person}{Susan Athey}, \bibinfo{person}{Filiz Garip},
  \bibinfo{person}{Thomas~L Griffiths}, \bibinfo{person}{Jon Kleinberg},
  \bibinfo{person}{Helen Margetts}, \bibinfo{person}{Sendhil Mullainathan},
  \bibinfo{person}{Matthew~J Salganik}, \bibinfo{person}{Simine Vazire},
  {et~al\mbox{.}}} \bibinfo{year}{2021}\natexlab{}.
\newblock \showarticletitle{Integrating explanation and prediction in
  computational social science}.
\newblock \bibinfo{journal}{\emph{Nature}} \bibinfo{volume}{595},
  \bibinfo{number}{7866} (\bibinfo{year}{2021}).
\newblock


\bibitem[Horn et~al\mbox{.}(2021)]%
        {horn2021investigating}
\bibfield{author}{\bibinfo{person}{Abigail~L Horn}, \bibinfo{person}{Brooke~M
  Bell}, \bibinfo{person}{Bernardo Garcia~Bulle Bueno}, \bibinfo{person}{Mohsen
  Bahrami}, \bibinfo{person}{Burcin Bozkaya}, \bibinfo{person}{Yan Cui},
  \bibinfo{person}{John~P Wilson}, \bibinfo{person}{Alex Pentland},
  \bibinfo{person}{Esteban~Moro Egido}, {and} \bibinfo{person}{Kayla de~la
  Haye}.} \bibinfo{year}{2021}\natexlab{}.
\newblock \showarticletitle{Investigating mobility-based fast food outlet
  visits as indicators of dietary intake and diet-related disease}.
\newblock \bibinfo{journal}{\emph{medRxiv}} (\bibinfo{year}{2021}).
\newblock


\bibitem[Howell et~al\mbox{.}(2016)]%
        {Howell:2016:ATP:2896338.2896358}
\bibfield{author}{\bibinfo{person}{Patrick~D. Howell},
  \bibinfo{person}{Layla~D. Martin}, \bibinfo{person}{Hesamoddin Salehian},
  \bibinfo{person}{Chul Lee}, \bibinfo{person}{Kyler~M. Eastman}, {and}
  \bibinfo{person}{Joohyun Kim}.} \bibinfo{year}{2016}\natexlab{}.
\newblock \showarticletitle{Analyzing taste preferences from crowdsourced food
  entries}. In \bibinfo{booktitle}{\emph{Proc. of the 6th International
  Conference on Digital Health Conference (DH)}}.
\newblock


\bibitem[Iwata et~al\mbox{.}(2009)]%
        {iwata2009topic}
\bibfield{author}{\bibinfo{person}{Tomoharu Iwata}, \bibinfo{person}{Shinji
  Watanabe}, \bibinfo{person}{Takeshi Yamada}, {and} \bibinfo{person}{Naonori
  Ueda}.} \bibinfo{year}{2009}\natexlab{}.
\newblock \showarticletitle{Topic tracking model for analyzing consumer
  purchase behavior}. In \bibinfo{booktitle}{\emph{Proc. Twenty-First
  International Joint Conference on Artificial Intelligence (IJCAI)}}.
\newblock


\bibitem[Kawamae(2010)]%
        {kawamae2010serendipitous}
\bibfield{author}{\bibinfo{person}{Noriaki Kawamae}.}
  \bibinfo{year}{2010}\natexlab{}.
\newblock \showarticletitle{Serendipitous recommendations via innovators}. In
  \bibinfo{booktitle}{\emph{Proc. of the 33rd International ACM Conference on
  Research and Development in Information Retrieval (SIGIR)}}.
\newblock


\bibitem[Kim et~al\mbox{.}(2021)]%
        {kim2021systematic}
\bibfield{author}{\bibinfo{person}{Jiin Kim}, \bibinfo{person}{Zara Ahmad},
  \bibinfo{person}{Yena Lee}, \bibinfo{person}{Flora Nasri},
  \bibinfo{person}{Hartej Gill}, \bibinfo{person}{Roger Mclntyre},
  \bibinfo{person}{Lee Phan}, {and} \bibinfo{person}{Leanna Lui}.}
  \bibinfo{year}{2021}\natexlab{}.
\newblock \showarticletitle{Systematic Review of the Validity of Screening
  Depression through Facebook, Twitter, and Instagram}.
\newblock \bibinfo{journal}{\emph{Journal of Affective Disorders}}
  (\bibinfo{year}{2021}).
\newblock


\bibitem[Kulshrestha et~al\mbox{.}(2019)]%
        {kulshrestha2019search}
\bibfield{author}{\bibinfo{person}{Juhi Kulshrestha},
  \bibinfo{person}{Motahhare Eslami}, \bibinfo{person}{Johnnatan Messias},
  \bibinfo{person}{Muhammad~Bilal Zafar}, \bibinfo{person}{Saptarshi Ghosh},
  \bibinfo{person}{Krishna~P Gummadi}, {and} \bibinfo{person}{Karrie
  Karahalios}.} \bibinfo{year}{2019}\natexlab{}.
\newblock \showarticletitle{Search bias quantification: Investigating political
  bias in social media and web search}.
\newblock \bibinfo{journal}{\emph{Information Retrieval Journal}}
  \bibinfo{volume}{22}, \bibinfo{number}{1} (\bibinfo{year}{2019}).
\newblock


\bibitem[Lazer et~al\mbox{.}(2021)]%
        {lazer2021meaningful}
\bibfield{author}{\bibinfo{person}{David Lazer}, \bibinfo{person}{Eszter
  Hargittai}, \bibinfo{person}{Deen Freelon}, \bibinfo{person}{Sandra
  Gonzalez-Bailon}, \bibinfo{person}{Kevin Munger}, \bibinfo{person}{Katherine
  Ognyanova}, {and} \bibinfo{person}{Jason Radford}.}
  \bibinfo{year}{2021}\natexlab{}.
\newblock \showarticletitle{Meaningful measures of human society in the
  twenty-first century}.
\newblock \bibinfo{journal}{\emph{Nature}} \bibinfo{volume}{595},
  \bibinfo{number}{7866} (\bibinfo{year}{2021}).
\newblock


\bibitem[Lazer et~al\mbox{.}(2014)]%
        {lazer2014parable}
\bibfield{author}{\bibinfo{person}{David Lazer}, \bibinfo{person}{Ryan
  Kennedy}, \bibinfo{person}{Gary King}, {and} \bibinfo{person}{Alessandro
  Vespignani}.} \bibinfo{year}{2014}\natexlab{}.
\newblock \showarticletitle{The parable of Google Flu: traps in big data
  analysis}.
\newblock \bibinfo{journal}{\emph{Science}} \bibinfo{volume}{343},
  \bibinfo{number}{6176} (\bibinfo{year}{2014}).
\newblock


\bibitem[Luo et~al\mbox{.}(2019)]%
        {luo2019co}
\bibfield{author}{\bibinfo{person}{Yuhan Luo}, \bibinfo{person}{Peiyi Liu},
  {and} \bibinfo{person}{Eun~Kyoung Choe}.} \bibinfo{year}{2019}\natexlab{}.
\newblock \showarticletitle{Co-Designing food trackers with dietitians:
  Identifying design opportunities for food tracker customization}. In
  \bibinfo{booktitle}{\emph{Proceedings of the 2019 Conference on Human Factors
  in Computing Systems (CHI)}}.
\newblock


\bibitem[Malik et~al\mbox{.}(2015)]%
        {malik2015population}
\bibfield{author}{\bibinfo{person}{Momin~M Malik}, \bibinfo{person}{Hemank
  Lamba}, \bibinfo{person}{Constantine Nakos}, {and}
  \bibinfo{person}{J{\"u}rgen Pfeffer}.} \bibinfo{year}{2015}\natexlab{}.
\newblock \showarticletitle{Population bias in geotagged tweets}. In
  \bibinfo{booktitle}{\emph{Proc. of the 9th International AAAI Conference on
  Web and Social Media (ICWSM)}}.
\newblock


\bibitem[Malik and Pfeffer(2016)]%
        {malik2016identifying}
\bibfield{author}{\bibinfo{person}{Momin~M Malik} {and}
  \bibinfo{person}{J{\"u}rgen Pfeffer}.} \bibinfo{year}{2016}\natexlab{}.
\newblock \showarticletitle{Identifying platform effects in social media data}.
  In \bibinfo{booktitle}{\emph{Proc. of the 10th International AAAI Conference
  on Web and Social Media (ICWSM)}}.
\newblock


\bibitem[Maystre and Grossglauser(2015)]%
        {maystre2015fast}
\bibfield{author}{\bibinfo{person}{Lucas Maystre} {and}
  \bibinfo{person}{Matthias Grossglauser}.} \bibinfo{year}{2015}\natexlab{}.
\newblock \showarticletitle{Fast and accurate inference of Plackett--Luce
  models}. In \bibinfo{booktitle}{\emph{Advances in Neural Information
  Processing Systems (NeurIPS)}}.
\newblock


\bibitem[Mejova et~al\mbox{.}(2016)]%
        {mejova2016fetishizing}
\bibfield{author}{\bibinfo{person}{Yelena Mejova}, \bibinfo{person}{Sofiane
  Abbar}, {and} \bibinfo{person}{Hamed Haddadi}.}
  \bibinfo{year}{2016}\natexlab{}.
\newblock \showarticletitle{Fetishizing food in digital age:\# foodporn around
  the world}. In \bibinfo{booktitle}{\emph{Proc. of the 10th International AAAI
  Conference on Web and Social Media (ICWSM)}}.
\newblock


\bibitem[Mejova et~al\mbox{.}(2015a)]%
        {mejova2015dietary}
\bibfield{author}{\bibinfo{person}{Yelena Mejova}, \bibinfo{person}{Hamed
  Haddadi}, \bibinfo{person}{Sofiane Abbar}, \bibinfo{person}{Azadeh
  Ghahghaei}, {and} \bibinfo{person}{Ingmar Weber}.}
  \bibinfo{year}{2015}\natexlab{a}.
\newblock \showarticletitle{Dietary habits of an expat nation: Case of Qatar}.
  In \bibinfo{booktitle}{\emph{2015 International Conference on Healthcare
  Informatics}}.
\newblock


\bibitem[Mejova et~al\mbox{.}(2015b)]%
        {mejova2015foodporn}
\bibfield{author}{\bibinfo{person}{Yelena Mejova}, \bibinfo{person}{Hamed
  Haddadi}, \bibinfo{person}{Anastasios Noulas}, {and} \bibinfo{person}{Ingmar
  Weber}.} \bibinfo{year}{2015}\natexlab{b}.
\newblock \showarticletitle{\# foodporn: Obesity patterns in culinary
  interactions}. In \bibinfo{booktitle}{\emph{Proc. of the 5th International
  Conference on Digital Health 2015}}.
\newblock


\bibitem[Mejova et~al\mbox{.}(2015c)]%
        {mejova2015twitter}
\bibfield{author}{\bibinfo{person}{Yelena Mejova}, \bibinfo{person}{Ingmar
  Weber}, {and} \bibinfo{person}{Michael~W Macy}.}
  \bibinfo{year}{2015}\natexlab{c}.
\newblock \bibinfo{booktitle}{\emph{Twitter: A digital socioscope}}.
\newblock


\bibitem[Mete et~al\mbox{.}(2019)]%
        {mete2019healthy}
\bibfield{author}{\bibinfo{person}{Rebecca Mete}, \bibinfo{person}{Alison
  Shield}, \bibinfo{person}{Kristen Murray}, \bibinfo{person}{Rachel Bacon},
  {and} \bibinfo{person}{Jane Kellett}.} \bibinfo{year}{2019}\natexlab{}.
\newblock \showarticletitle{What is healthy eating? A qualitative exploration}.
\newblock \bibinfo{journal}{\emph{Public Health Nutrition}}
  \bibinfo{volume}{22}, \bibinfo{number}{13} (\bibinfo{year}{2019}).
\newblock


\bibitem[Meyre(2017)]%
        {meyre2017agriculture}
\bibfield{author}{\bibinfo{person}{S Meyre}.} \bibinfo{year}{2017}\natexlab{}.
\newblock \bibinfo{title}{Agriculture et alimentation. Statistique de poche
  2017}.
\newblock
\newblock


\bibitem[Mohanty et~al\mbox{.}(2022)]%
        {mohanty2021food}
\bibfield{author}{\bibinfo{person}{Sharada~Prasanna Mohanty},
  \bibinfo{person}{Gaurav Singhal}, \bibinfo{person}{Eric~Antoine Scuccimarra},
  \bibinfo{person}{Djilani Kebaili}, \bibinfo{person}{Harris H{\'e}ritier},
  \bibinfo{person}{Victor Boulanger}, {and} \bibinfo{person}{Marcel
  Salath{\'e}}.} \bibinfo{year}{2022}\natexlab{}.
\newblock \showarticletitle{The Food Recognition Benchmark: Using Deep Learning
  to Recognize Food in Images}.
\newblock \bibinfo{journal}{\emph{Frontiers in Nutrition}}  \bibinfo{volume}{9}
  (\bibinfo{year}{2022}).
\newblock


\bibitem[Morstatter et~al\mbox{.}(2014)]%
        {morstatter2014biased}
\bibfield{author}{\bibinfo{person}{Fred Morstatter},
  \bibinfo{person}{J{\"u}rgen Pfeffer}, {and} \bibinfo{person}{Huan Liu}.}
  \bibinfo{year}{2014}\natexlab{}.
\newblock \showarticletitle{When is it biased? Assessing the representativeness
  of Twitter's streaming API}. In \bibinfo{booktitle}{\emph{Proc. of the 23rd
  International Conference on World Wide Web (TheWebConf)}}.
\newblock


\bibitem[Naritomi and Yanai(2020)]%
        {naritomi2020caloriecaptorglass}
\bibfield{author}{\bibinfo{person}{Shu Naritomi} {and} \bibinfo{person}{Keiji
  Yanai}.} \bibinfo{year}{2020}\natexlab{}.
\newblock \showarticletitle{CalorieCaptorGlass: Food calorie estimation based
  on actual size using hololens and deep learning}. In
  \bibinfo{booktitle}{\emph{IEEE Conference on Virtual Reality and 3D User
  Interfaces Abstracts and Workshops (VRW)}}.
\newblock


\bibitem[Ofli et~al\mbox{.}(2017)]%
        {ofli2017saki}
\bibfield{author}{\bibinfo{person}{Ferda Ofli}, \bibinfo{person}{Yusuf Aytar},
  \bibinfo{person}{Ingmar Weber}, \bibinfo{person}{Raggi Al~Hammouri}, {and}
  \bibinfo{person}{Antonio Torralba}.} \bibinfo{year}{2017}\natexlab{}.
\newblock \showarticletitle{Is saki\# delicious?: The food perception gap on
  Instagram and its relation to health}. In \bibinfo{booktitle}{\emph{Proc. of
  the 26th International Conference on World Wide Web (TheWebConf)}}.
\newblock


\bibitem[Okamoto and Yanai(2021)]%
        {okamoto2021uec}
\bibfield{author}{\bibinfo{person}{Kaimu Okamoto} {and} \bibinfo{person}{Keiji
  Yanai}.} \bibinfo{year}{2021}\natexlab{}.
\newblock \showarticletitle{UEC-FoodPIX Complete: A Large-scale Food Image
  Segmentation Dataset}. In \bibinfo{booktitle}{\emph{International Conference
  on Pattern Recognition}}.
\newblock


\bibitem[Olteanu et~al\mbox{.}(2019)]%
        {olteanu2019social}
\bibfield{author}{\bibinfo{person}{Alexandra Olteanu}, \bibinfo{person}{Carlos
  Castillo}, \bibinfo{person}{Fernando Diaz}, {and} \bibinfo{person}{Emre
  K{\i}c{\i}man}.} \bibinfo{year}{2019}\natexlab{}.
\newblock \showarticletitle{Social data: Biases, methodological pitfalls, and
  ethical boundaries}.
\newblock \bibinfo{journal}{\emph{Frontiers in Big Data}}  \bibinfo{volume}{2}
  (\bibinfo{year}{2019}).
\newblock


\bibitem[Pater et~al\mbox{.}(2016)]%
        {hunger2016}
\bibfield{author}{\bibinfo{person}{Jessica~A. Pater},
  \bibinfo{person}{Oliver~L. Haimson}, \bibinfo{person}{Nazanin Andalibi},
  {and} \bibinfo{person}{Elizabeth~D. Mynatt}.}
  \bibinfo{year}{2016}\natexlab{}.
\newblock \showarticletitle{``Hunger hurts but starving works'': Characterizing
  the presentation of eating disorders online}. In
  \bibinfo{booktitle}{\emph{Proc. of the 2016 ACM Conference on Computer
  Supported Cooperative Work and Social Computing (CSCW)}}.
\newblock


\bibitem[Pellert et~al\mbox{.}(2022)]%
        {pellert2022validating}
\bibfield{author}{\bibinfo{person}{Max Pellert}, \bibinfo{person}{Hannah
  Metzler}, \bibinfo{person}{Michael Matzenberger}, {and}
  \bibinfo{person}{David Garcia}.} \bibinfo{year}{2022}\natexlab{}.
\newblock \showarticletitle{Validating daily social media macroscopes of
  emotions}.
\newblock \bibinfo{journal}{\emph{Scientific Reports}} \bibinfo{volume}{12},
  \bibinfo{number}{1} (\bibinfo{year}{2022}).
\newblock


\bibitem[Perez-Ortiz and Mantiuk(2017)]%
        {perez2017practical}
\bibfield{author}{\bibinfo{person}{Maria Perez-Ortiz} {and}
  \bibinfo{person}{Rafal~K Mantiuk}.} \bibinfo{year}{2017}\natexlab{}.
\newblock \showarticletitle{A practical guide and software for analysing
  pairwise comparison experiments}.
\newblock \bibinfo{journal}{\emph{arXiv preprint arXiv:1712.03686}}
  (\bibinfo{year}{2017}).
\newblock


\bibitem[Phan and Gatica-Perez(2017)]%
        {phan2017healthy}
\bibfield{author}{\bibinfo{person}{Thanh-Trung Phan} {and}
  \bibinfo{person}{Daniel Gatica-Perez}.} \bibinfo{year}{2017}\natexlab{}.
\newblock \showarticletitle{Healthy\# fondue\# dinner: analysis and inference
  of food and drink consumption patterns on instagram}. In
  \bibinfo{booktitle}{\emph{Proc. of the 16th International Conference on
  Mobile and Ubiquitous Multimedia}}.
\newblock


\bibitem[Poushter et~al\mbox{.}(2016)]%
        {poushter2016smartphone}
\bibfield{author}{\bibinfo{person}{Jacob Poushter} {et~al\mbox{.}}}
  \bibinfo{year}{2016}\natexlab{}.
\newblock \showarticletitle{Smartphone ownership and internet usage continues
  to climb in emerging economies}.
\newblock \bibinfo{journal}{\emph{Pew research center}} \bibinfo{volume}{22},
  \bibinfo{number}{1} (\bibinfo{year}{2016}).
\newblock


\bibitem[Ribeiro et~al\mbox{.}(2021)]%
        {ribeiro2021sudden}
\bibfield{author}{\bibinfo{person}{Manoel~Horta Ribeiro},
  \bibinfo{person}{Kristina Gligori{\'c}}, \bibinfo{person}{Maxime Peyrard},
  \bibinfo{person}{Florian Lemmerich}, \bibinfo{person}{Markus Strohmaier},
  {and} \bibinfo{person}{Robert West}.} \bibinfo{year}{2021}\natexlab{}.
\newblock \showarticletitle{Sudden attention shifts on Wikipedia during the
  COVID-19 crisis}. In \bibinfo{booktitle}{\emph{Proc. of the 15th
  International AAAI Conference on Web and Social Media (ICWSM)}}.
\newblock


\bibitem[Rokicki et~al\mbox{.}(2018)]%
        {rokicki2018impact}
\bibfield{author}{\bibinfo{person}{Markus Rokicki}, \bibinfo{person}{Christoph
  Trattner}, {and} \bibinfo{person}{Eelco Herder}.}
  \bibinfo{year}{2018}\natexlab{}.
\newblock \showarticletitle{The impact of recipe features, social cues and
  demographics on estimating the healthiness of online recipes}. In
  \bibinfo{booktitle}{\emph{Proc. of the 12th International AAAI Conference on
  Web and Social Media (ICWSM)}}.
\newblock


\bibitem[Ruths and Pfeffer(2014)]%
        {ruths2014social}
\bibfield{author}{\bibinfo{person}{Derek Ruths} {and}
  \bibinfo{person}{J{\"u}rgen Pfeffer}.} \bibinfo{year}{2014}\natexlab{}.
\newblock \showarticletitle{Social media for large studies of behavior}.
\newblock \bibinfo{journal}{\emph{Science}} \bibinfo{volume}{346},
  \bibinfo{number}{6213} (\bibinfo{year}{2014}).
\newblock


\bibitem[Sadilek et~al\mbox{.}(2018)]%
        {sadilek2018machine}
\bibfield{author}{\bibinfo{person}{Adam Sadilek}, \bibinfo{person}{Stephanie
  Caty}, \bibinfo{person}{Lauren DiPrete}, \bibinfo{person}{Raed Mansour},
  \bibinfo{person}{Tom Schenk}, \bibinfo{person}{Mark Bergtholdt},
  \bibinfo{person}{Ashish Jha}, \bibinfo{person}{Prem Ramaswami}, {and}
  \bibinfo{person}{Evgeniy Gabrilovich}.} \bibinfo{year}{2018}\natexlab{}.
\newblock \showarticletitle{Machine-learned epidemiology: real-time detection
  of foodborne illness at scale}.
\newblock \bibinfo{journal}{\emph{npj Digital Medicine}} \bibinfo{volume}{1},
  \bibinfo{number}{1} (\bibinfo{year}{2018}).
\newblock


\bibitem[Saha et~al\mbox{.}(2021)]%
        {saha2021life}
\bibfield{author}{\bibinfo{person}{Koustuv Saha}, \bibinfo{person}{Jordyn
  Seybolt}, \bibinfo{person}{Stephen~M Mattingly}, \bibinfo{person}{Talayeh
  Aledavood}, \bibinfo{person}{Chaitanya Konjeti}, \bibinfo{person}{Gonzalo~J
  Martinez}, \bibinfo{person}{Ted Grover}, \bibinfo{person}{Gloria Mark}, {and}
  \bibinfo{person}{Munmun De~Choudhury}.} \bibinfo{year}{2021}\natexlab{}.
\newblock \showarticletitle{What Life Events are Disclosed on Social Media,
  How, When, and By Whom?}. In \bibinfo{booktitle}{\emph{Proc. of the 2021
  Conference on Human Factors in Computing Systems (CHI)}}.
\newblock


\bibitem[Sahoo et~al\mbox{.}(2019)]%
        {sahoo2019foodai}
\bibfield{author}{\bibinfo{person}{Doyen Sahoo}, \bibinfo{person}{Wang Hao},
  \bibinfo{person}{Shu Ke}, \bibinfo{person}{Wu Xiongwei},
  \bibinfo{person}{Hung Le}, \bibinfo{person}{Palakorn Achananuparp},
  \bibinfo{person}{Ee-Peng Lim}, {and} \bibinfo{person}{Steven~CH Hoi}.}
  \bibinfo{year}{2019}\natexlab{}.
\newblock \showarticletitle{FoodAI: food image recognition via deep learning
  for smart food logging}. In \bibinfo{booktitle}{\emph{Proc. of the 25th ACM
  International Conference on Knowledge Discovery \& Data Mining (KDD)}}.
\newblock


\bibitem[Sajadmanesh et~al\mbox{.}(2017)]%
        {sajadmanesh2017kissing}
\bibfield{author}{\bibinfo{person}{Sina Sajadmanesh}, \bibinfo{person}{Sina
  Jafarzadeh}, \bibinfo{person}{Seyed~Ali Ossia}, \bibinfo{person}{Hamid~R
  Rabiee}, \bibinfo{person}{Hamed Haddadi}, \bibinfo{person}{Yelena Mejova},
  \bibinfo{person}{Mirco Musolesi}, \bibinfo{person}{Emiliano~De Cristofaro},
  {and} \bibinfo{person}{Gianluca Stringhini}.}
  \bibinfo{year}{2017}\natexlab{}.
\newblock \showarticletitle{Kissing cuisines: Exploring worldwide culinary
  habits on the web}. In \bibinfo{booktitle}{\emph{Proc. of the 26th
  International Conference on World Wide Web (TheWebConf)}}.
\newblock


\bibitem[Salganik(2019)]%
        {salganik2019bit}
\bibfield{author}{\bibinfo{person}{Matthew~J Salganik}.}
  \bibinfo{year}{2019}\natexlab{}.
\newblock \bibinfo{booktitle}{\emph{Bit by bit: Social research in the digital
  age}}.
\newblock


\bibitem[Salvador et~al\mbox{.}(2017)]%
        {salvador2017learning}
\bibfield{author}{\bibinfo{person}{Amaia Salvador}, \bibinfo{person}{Nicholas
  Hynes}, \bibinfo{person}{Yusuf Aytar}, \bibinfo{person}{Javier Marin},
  \bibinfo{person}{Ferda Ofli}, \bibinfo{person}{Ingmar Weber}, {and}
  \bibinfo{person}{Antonio Torralba}.} \bibinfo{year}{2017}\natexlab{}.
\newblock \showarticletitle{Learning cross-modal embeddings for cooking recipes
  and food images}. In \bibinfo{booktitle}{\emph{Proc. of the IEEE Conference
  on Computer Vision and Pattern Recognition (CVPR)}}.
\newblock


\bibitem[Saura et~al\mbox{.}(2020)]%
        {saura2020gaining}
\bibfield{author}{\bibinfo{person}{Jose~Ramon Saura}, \bibinfo{person}{Ana
  Reyes-Menendez}, {and} \bibinfo{person}{Stephen~B Thomas}.}
  \bibinfo{year}{2020}\natexlab{}.
\newblock \showarticletitle{Gaining a deeper understanding of nutrition using
  social networks and user-generated content}.
\newblock \bibinfo{journal}{\emph{Internet Interventions}}
  \bibinfo{volume}{20} (\bibinfo{year}{2020}).
\newblock


\bibitem[Schroeder et~al\mbox{.}(2017)]%
        {schroeder2017supporting}
\bibfield{author}{\bibinfo{person}{Jessica Schroeder}, \bibinfo{person}{Jane
  Hoffswell}, \bibinfo{person}{Chia-Fang Chung}, \bibinfo{person}{James
  Fogarty}, \bibinfo{person}{Sean Munson}, {and} \bibinfo{person}{Jasmine
  Zia}.} \bibinfo{year}{2017}\natexlab{}.
\newblock \showarticletitle{Supporting patient-provider collaboration to
  identify individual triggers using food and symptom journals}. In
  \bibinfo{booktitle}{\emph{Proc. of the 2017 ACM Conference on Computer
  Supported Cooperative Work and Social Computing (CSCW)}}.
\newblock


\bibitem[Sen et~al\mbox{.}(2020)]%
        {sen2020reliability}
\bibfield{author}{\bibinfo{person}{Indira Sen}, \bibinfo{person}{Fabian
  Fl{\"o}ck}, {and} \bibinfo{person}{Claudia Wagner}.}
  \bibinfo{year}{2020}\natexlab{}.
\newblock \showarticletitle{On the reliability and validity of detecting
  approval of political actors in tweets}. In \bibinfo{booktitle}{\emph{Proc.
  of the 2020 Conference on Empirical Methods in Natural Language Processing
  (EMNLP)}}.
\newblock


\bibitem[Sen et~al\mbox{.}(2021)]%
        {sen2021total}
\bibfield{author}{\bibinfo{person}{Indira Sen}, \bibinfo{person}{Fabian
  Fl{\"o}ck}, \bibinfo{person}{Katrin Weller}, \bibinfo{person}{Bernd
  Wei{\ss}}, {and} \bibinfo{person}{Claudia Wagner}.}
  \bibinfo{year}{2021}\natexlab{}.
\newblock \showarticletitle{A total error framework for digital traces of human
  behavior on online platforms}.
\newblock \bibinfo{journal}{\emph{Public Opinion Quarterly}}
  (\bibinfo{year}{2021}).
\newblock


\bibitem[Sharma and De~Choudhury(2015)]%
        {sharma2015measuring}
\bibfield{author}{\bibinfo{person}{Sanket~S Sharma} {and}
  \bibinfo{person}{Munmun De~Choudhury}.} \bibinfo{year}{2015}\natexlab{}.
\newblock \showarticletitle{Measuring and characterizing nutritional
  information of food and ingestion content in Instagram}. In
  \bibinfo{booktitle}{\emph{Proc. of the 24th International Conference on World
  Wide Web (TheWebConf)}}.
\newblock


\bibitem[Singla et~al\mbox{.}(2016)]%
        {singla2016food}
\bibfield{author}{\bibinfo{person}{Ashutosh Singla}, \bibinfo{person}{Lin
  Yuan}, {and} \bibinfo{person}{Touradj Ebrahimi}.}
  \bibinfo{year}{2016}\natexlab{}.
\newblock \showarticletitle{Food/non-food image classification and food
  categorization using pre-trained googlenet model}. In
  \bibinfo{booktitle}{\emph{Proceedings of the 2nd International Workshop on
  Multimedia Assisted Dietary Management}}.
\newblock


\bibitem[Tobey et~al\mbox{.}(2019)]%
        {tobey2019low}
\bibfield{author}{\bibinfo{person}{Lauren~N Tobey}, \bibinfo{person}{Christine
  Mouzong}, \bibinfo{person}{Joyce~Senior Angulo}, \bibinfo{person}{Sally
  Bowman}, {and} \bibinfo{person}{Melinda~M Manore}.}
  \bibinfo{year}{2019}\natexlab{}.
\newblock \showarticletitle{How low-income mothers select and adapt recipes and
  implications for promoting healthy recipes online}.
\newblock \bibinfo{journal}{\emph{Nutrients}} \bibinfo{volume}{11},
  \bibinfo{number}{2} (\bibinfo{year}{2019}).
\newblock


\bibitem[Trattner et~al\mbox{.}(2018)]%
        {trattner2018predictability}
\bibfield{author}{\bibinfo{person}{Christoph Trattner},
  \bibinfo{person}{Dominik Moesslang}, {and} \bibinfo{person}{David
  Elsweiler}.} \bibinfo{year}{2018}\natexlab{}.
\newblock \showarticletitle{On the predictability of the popularity of online
  recipes}.
\newblock \bibinfo{journal}{\emph{EPJ Data Science}} \bibinfo{volume}{7},
  \bibinfo{number}{1} (\bibinfo{year}{2018}).
\newblock


\bibitem[Vosen and Schmidt(2011)]%
        {vosen2011forecasting}
\bibfield{author}{\bibinfo{person}{Simeon Vosen} {and} \bibinfo{person}{Torsten
  Schmidt}.} \bibinfo{year}{2011}\natexlab{}.
\newblock \showarticletitle{Forecasting private consumption: survey-based
  indicators vs. Google trends}.
\newblock \bibinfo{journal}{\emph{Journal of Forecasting}}
  \bibinfo{volume}{30}, \bibinfo{number}{6} (\bibinfo{year}{2011}).
\newblock


\bibitem[Wagner et~al\mbox{.}(2014a)]%
        {wagner2014nature}
\bibfield{author}{\bibinfo{person}{Claudia Wagner}, \bibinfo{person}{Philipp
  Singer}, {and} \bibinfo{person}{Markus Strohmaier}.}
  \bibinfo{year}{2014}\natexlab{a}.
\newblock \showarticletitle{The nature and evolution of online food
  preferences}.
\newblock \bibinfo{journal}{\emph{EPJ Data Science}}  \bibinfo{volume}{3}
  (\bibinfo{year}{2014}).
\newblock


\bibitem[Wagner et~al\mbox{.}(2014b)]%
        {wagner2014spatial}
\bibfield{author}{\bibinfo{person}{Claudia Wagner}, \bibinfo{person}{Philipp
  Singer}, {and} \bibinfo{person}{Markus Strohmaier}.}
  \bibinfo{year}{2014}\natexlab{b}.
\newblock \showarticletitle{Spatial and temporal patterns of online food
  preferences}. In \bibinfo{booktitle}{\emph{Proc. of the 23rd International
  Conference on World Wide Web (TheWebConf)}}.
\newblock


\bibitem[Wagner et~al\mbox{.}(2021)]%
        {wagner2021measuring}
\bibfield{author}{\bibinfo{person}{Claudia Wagner}, \bibinfo{person}{Markus
  Strohmaier}, \bibinfo{person}{Alexandra Olteanu}, \bibinfo{person}{Emre
  K{\i}c{\i}man}, \bibinfo{person}{Noshir Contractor}, {and}
  \bibinfo{person}{Tina Eliassi-Rad}.} \bibinfo{year}{2021}\natexlab{}.
\newblock \showarticletitle{Measuring algorithmically infused societies}.
\newblock \bibinfo{journal}{\emph{Nature}} \bibinfo{volume}{595},
  \bibinfo{number}{7866} (\bibinfo{year}{2021}).
\newblock


\bibitem[Wang et~al\mbox{.}(2015)]%
        {wang2015forecasting}
\bibfield{author}{\bibinfo{person}{Wei Wang}, \bibinfo{person}{David
  Rothschild}, \bibinfo{person}{Sharad Goel}, {and} \bibinfo{person}{Andrew
  Gelman}.} \bibinfo{year}{2015}\natexlab{}.
\newblock \showarticletitle{Forecasting elections with non-representative
  polls}.
\newblock \bibinfo{journal}{\emph{International Journal of Forecasting}}
  \bibinfo{volume}{31}, \bibinfo{number}{3} (\bibinfo{year}{2015}).
\newblock


\bibitem[Weber et~al\mbox{.}(2006)]%
        {weber2006internet}
\bibfield{author}{\bibinfo{person}{Kristi Weber}, \bibinfo{person}{Mary Story},
  {and} \bibinfo{person}{Lisa Harnack}.} \bibinfo{year}{2006}\natexlab{}.
\newblock \showarticletitle{Internet food marketing strategies aimed at
  children and adolescents: A content analysis of food and beverage brand web
  sites}.
\newblock \bibinfo{journal}{\emph{Journal of the American Dietetic
  Association}} \bibinfo{number}{9} (\bibinfo{year}{2006}).
\newblock


\bibitem[West et~al\mbox{.}(2013)]%
        {West:2013:CCI:2488388.2488510}
\bibfield{author}{\bibinfo{person}{Robert West}, \bibinfo{person}{Ryen~W.
  White}, {and} \bibinfo{person}{Eric Horvitz}.}
  \bibinfo{year}{2013}\natexlab{}.
\newblock \showarticletitle{From Cookies to Cooks: Insights on Dietary Patterns
  via Analysis of Web Usage Logs}. In \bibinfo{booktitle}{\emph{Proc. of the
  22nd International Conference on World Wide Web (TheWebConf)}}.
\newblock


\bibitem[Widener and Li(2014)]%
        {widener2014using}
\bibfield{author}{\bibinfo{person}{Michael~J Widener} {and}
  \bibinfo{person}{Wenwen Li}.} \bibinfo{year}{2014}\natexlab{}.
\newblock \showarticletitle{Using geolocated Twitter data to monitor the
  prevalence of healthy and unhealthy food references across the US}.
\newblock \bibinfo{journal}{\emph{Applied Geography}}  \bibinfo{volume}{54}
  (\bibinfo{year}{2014}).
\newblock


\bibitem[Wouters et~al\mbox{.}(2010)]%
        {wouters2010peer}
\bibfield{author}{\bibinfo{person}{Eveline~J Wouters},
  \bibinfo{person}{Junilla~K Larsen}, \bibinfo{person}{Stef~P Kremers},
  \bibinfo{person}{Pieter~C Dagnelie}, {and} \bibinfo{person}{Rinie Geenen}.}
  \bibinfo{year}{2010}\natexlab{}.
\newblock \showarticletitle{Peer influence on snacking behavior in
  adolescence}.
\newblock \bibinfo{journal}{\emph{Appetite}} \bibinfo{volume}{55},
  \bibinfo{number}{1} (\bibinfo{year}{2010}).
\newblock


\bibitem[Yanai and Kawano(2014)]%
        {yanai2014twitter}
\bibfield{author}{\bibinfo{person}{Keiji Yanai} {and}
  \bibinfo{person}{Yoshiyuki Kawano}.} \bibinfo{year}{2014}\natexlab{}.
\newblock \showarticletitle{Twitter food photo mining and analysis for one
  hundred kinds of foods}. In \bibinfo{booktitle}{\emph{Pacific Rim Conference
  on Multimedia}}.
\newblock


\end{thebibliography}

\received{January 2022}
\received[revised]{April 2022}
\received[accepted]{August 2022}




\end{document}
\endinput